\documentclass[9pt,conference,compsoc]{IEEEtran}
\IEEEoverridecommandlockouts

\usepackage{newpxtext,newpxmath}
\usepackage{float}
\usepackage{caption}
\usepackage{algorithm}
\usepackage{graphicx}
\usepackage{tabularx} 
\usepackage{threeparttable}
\usepackage{amsmath}
\usepackage{placeins}
\usepackage[normalem]{ulem}
\usepackage{latexsym}
\usepackage{subcaption}
\usepackage{times}
\usepackage{enumerate}
\usepackage{xspace}
\usepackage{calc}
\usepackage{url}
\usepackage{booktabs}
\usepackage[english]{babel}
\usepackage{algorithm, algcompatible}
\usepackage[noend]{algpseudocode}
\usepackage{hhline}
\usepackage{multirow}
\usepackage{makecell} 
\usepackage{hyperref}
\hypersetup{
	citecolor=green
}
\usepackage{wrapfig}
\usepackage{csquotes}
\usepackage{multicol}
\usepackage{tikz}
\usepackage{xparse}
\usepackage{orcidlink}
\usepackage{etoolbox}
\usepackage{amsmath,amsfonts}
\usepackage{graphicx}
\usepackage{textcomp}
\usepackage{xcolor}

\usepackage[backend=bibtex,sorting=none,style=ieee,url=false]{biblatex}

\usepackage{tikz}
\usetikzlibrary{positioning, arrows.meta, shapes.misc}

\newcommand{\sgn}{\ensuremath {\texttt{SGN}}{\xspace}}
\newcommand{\sgnkg}{\ensuremath {\texttt{SGN.Kg}}{\xspace}}
\newcommand{\sgnsig}{\ensuremath {\texttt{SGN.Sig}}{\xspace}}
\newcommand{\sgnver}{\ensuremath {\texttt{SGN.Ver}}{\xspace}}

\newcommand{\fsgn}{\ensuremath {\texttt{FSGN}}{\xspace}}
\newcommand{\fsgnkg}{\ensuremath {\texttt{FSGN.Kg}}{\xspace}}
\newcommand{\fsgnupd}{\ensuremath {\texttt{FSGN.Upd}}{\xspace}}
\newcommand{\fsgnsig}{\ensuremath {\texttt{FSGN.Sig}}{\xspace}}
\newcommand{\fsgnver}{\ensuremath {\texttt{FSGN.Ver}}{\xspace}}

\newcommand{\hdsgn}{\ensuremath {\texttt{HDSGN}{\xspace}}}
\newcommand{\hdsgnkg}{\ensuremath {\texttt{HDSGN.Kg}}{\xspace}}
\newcommand{\hdsgnsig}{\ensuremath {\texttt{HDSGN.Sig}}{\xspace}}
\newcommand{\hdsgnver}{\ensuremath {\texttt{HDSGN.Ver}}{\xspace}}
\newcommand{\hdsgncomconstr}{\ensuremath {\texttt{HDSGN.ComC}}{\xspace}}

\newcommand{\fhdsgn}{\ensuremath {\texttt{FHDSGN}{\xspace}}}
\newcommand{\fhdsgnkg}{\ensuremath {\texttt{FHDSGN.Kg}}{\xspace}}
\newcommand{\fhdsgnsig}{\ensuremath {\texttt{FHDSGN.Sig}}{\xspace}}
\newcommand{\fhdsgnver}{\ensuremath {\texttt{FHDSGN.Ver}}{\xspace}}
\newcommand{\fhdsgncomconstr}{\ensuremath {\texttt{FHDSGN.ComC}}{\xspace}}
\newcommand{\fhdsgnupd}{\ensuremath {\texttt{FHDSGN.Upd}}{\xspace}}

\newcommand{\lrsha}{\ensuremath {\texttt{LRSHA}{\xspace}}}
\newcommand{\lrshakg}{\ensuremath {\texttt{LRSHA.Kg}}{\xspace}}
\newcommand{\lrshasig}{\ensuremath {\texttt{LRSHA.Sig}}{\xspace}}
\newcommand{\lrshaver}{\ensuremath {\texttt{LRSHA.Ver}}{\xspace}}
\newcommand{\lrshacomconstr}{\ensuremath {\texttt{LRSHA.ComC}}{\xspace}}

\newcommand{\flrsha}{\ensuremath {\texttt{FLRSHA}{\xspace}}}
\newcommand{\flrshakg}{\ensuremath {\texttt{FLRSHA.Kg}}{\xspace}}
\newcommand{\flrshasig}{\ensuremath {\texttt{FLRSHA.Sig}}{\xspace}}
\newcommand{\flrshaver}{\ensuremath {\texttt{FLRSHA.Ver}}{\xspace}}
\newcommand{\flrshacomconstr}{\ensuremath {\texttt{FLRSHA.ComC}}{\xspace}}
\newcommand{\flrshaupd}{\ensuremath {\texttt{FLRSHA.Upd}}{\xspace}}

\newcommand{\xmssmt}{\ensuremath {\text{XMSS}^\text{MT}{\xspace}}}

\newcommand{\kg}{\ensuremath {\texttt{Kg}}{\xspace}}
\newcommand{\upd}{\ensuremath {\texttt{Upd}}{\xspace}}
\newcommand{\ssig}{\ensuremath {\texttt{Sig}}{\xspace}}
\newcommand{\ver}{\ensuremath {\texttt{Ver}}{\xspace}}
\newcommand{\comconstr}{\ensuremath {\texttt{ComC}}{\xspace}}
\newcommand{\comc}{\ensuremath {\text{ComC}}{\xspace}}

\newcommand{\sk}{\ensuremath { \mathit{sk} }{\xspace}}
\newcommand{\pk}{\ensuremath { \mathit{PK} }{\xspace}}

\newcommand{\comconstral}{\ensuremath {\texttt{ComC}_{\vec{a}}(.)}{\xspace}}
\newcommand{\Breakin}{\ensuremath {\texttt{Break-In(.)}}{\xspace}}
\newcommand{\ro}{\ensuremath {\mathit{RO}(.)}{\xspace}}

\newcommand{\Sgnsig}{\ensuremath{\texttt{SGN.Sig}_{\sk}(.)}}

\newcommand{\Fsgnsigj}{\ensuremath{\texttt{FSGN.Sig}_{\sk_j}(.)}}

\newcommand{\Hdsgnsig}{\ensuremath{\texttt{HDSGN.Sig}_{\sk}(.)}}
\newcommand{\Hdsgncomconstr}{\ensuremath{\texttt{HDSGN.ComC}_{\vec{a}}(.)}}

\newcommand{\Lrshasig}{\ensuremath{\texttt{LRSHA.Sig}_{\sk}(.)}}
\newcommand{\Lrshacomconstr}{\ensuremath{\texttt{LRSHA.ComC}_{\vec{a}}(.)}}

\newcommand{\Fhdsgnsig}{\ensuremath{\texttt{FHDSGN.Sig}_{\sk_j}(.)}}
\newcommand{\Fhdsgncomconstr}{\ensuremath{\texttt{FHDSGN.ComC}_{\vec{a}}(.)}}

\newcommand{\Flrshasig}{\ensuremath{\texttt{FLRSHA.Sig}_{\sk_j}(.)}}
\newcommand{\Flrshacomconstr}{\ensuremath{\texttt{FLRSHA.ComC}_{\vec{a}}(.)}}

\newcommand{\EUCMA}{\ensuremath { \texttt{EU-CMA} }{\xspace}}
\newcommand{\FEUCMA}{\ensuremath { \texttt{F-EU-CMA} }{\xspace}}
\newcommand{\HDEUCMA}{\ensuremath { \texttt{HD-EU-CMA} }{\xspace}}
\newcommand{\FHDEUCMA}{\ensuremath { \texttt{FHD-EU-CMA} }{\xspace}}

\newcommand{\advsgnA}{\ensuremath {\mathit{Adv}_{\sgn}^{\EUCMA}(\mathcal{A})}{\xspace}}
\newcommand{\advsgn}{\ensuremath {\mathit{Adv}_{\sgn}^{\EUCMA}(t,q_H,q_s)}{\xspace}}
\newcommand{\advsgnp}{\ensuremath {\mathit{Adv}_{\sgn}^{\EUCMA}(t'',q_H,q_s)}{\xspace}}
\newcommand{\advsgnproof}{\ensuremath {\mathit{Adv}_{\sgn}^{\EUCMA}(t'',q_H,L\cdot q_s)}{\xspace}}

\newcommand{\advfsgnA}{\ensuremath {\mathit{Adv}_{\fsgn}^{\FEUCMA}(\mathcal{A})}{\xspace}}
\newcommand{\advfsgn}{\ensuremath {\mathit{Adv}_{\fsgn}^{\FEUCMA}(t,q_H,q_s,1)}{\xspace}}
\newcommand{\advfsgnproof}{\ensuremath {\mathit{Adv}_{\fsgn}^{\EUCMA}(t'',q_H,L\cdot q_s)}{\xspace}}

\newcommand{\advhdsgnA}{\ensuremath{\mathit{Adv}_{\mathit{\hdsgn}}^{\HDEUCMA}(\mathcal{A})}{\xspace}}
\newcommand{\advhdsgn}{\ensuremath {\mathit{Adv}_{\mathit{\hdsgn}}^{\HDEUCMA}(t,q_H, q_s)}{\xspace}}

\newcommand{\advfhdsgnA}{\ensuremath{\mathit{Adv}_{\mathit{\fhdsgn}}^{\FHDEUCMA}(\mathcal{A})}{\xspace}}
\newcommand{\advfhdsgn}{\ensuremath {\mathit{Adv}_{\mathit{\fhdsgn}}^{\FHDEUCMA}(t,q_H, q_s, 1)}{\xspace}}

\newcommand{\advlrsha}{\ensuremath {\mathit{Adv}_{\mathit{\lrsha}}^{\HDEUCMA}(t,q_H, q_s)}{\xspace}}
\newcommand{\advflrsha}{\ensuremath {\mathit{Adv}_{\mathit{\flrsha}}^{\FHDEUCMA}(t,q_H, q_s,1)}{\xspace}}

\newcommand{\advAdll}{\ensuremath {\mathit{Adv}_{\mathbb{G}, \alpha}^{\dl}(\mathcal{A})}{\xspace}}

\newcommand{\advdll}{\ensuremath {\mathit{Adv}_{\mathbb{G}, \alpha}^{\dl}(t')}{\xspace}}

\newcommand{\lh}{\ensuremath {\mathcal{LH}}{\xspace}}
\newcommand{\lm}{\ensuremath {\mathcal{LM}}{\xspace}}
\newcommand{\lr}{\ensuremath {\mathcal{LR}}{\xspace}}

\newcommand{\Areal}{\ensuremath {\overrightarrow{A}_{\mathit{real}}}{\xspace}}
\newcommand{\Asim}{\ensuremath {\overrightarrow{A}_{\mathit{sim}}}{\xspace}}
\newcommand{\Ra}{\ensuremath \stackrel{\$}{\leftarrow}{\xspace}}
\newcommand{\Rq}{\ensuremath \stackrel{\$}{\leftarrow}\mathbb{Z}_{q}^{*}{\xspace}}
\newcommand{\Rp}{\ensuremath \stackrel{\$}{\leftarrow}\mathbb{Z}_{p}^{*}{\xspace}}
\newcommand{\as}{\ensuremath {\leftarrow}{\xspace}}
\newcommand{\prf}{\ensuremath {\texttt{PRF}}{\xspace}}
\newcommand{\bigo}{\ensuremath{\mathcal{O}}}
\newcommand{\dlp}{\ensuremath { \texttt{DLP} }{\xspace}}
\newcommand{\dl}{\ensuremath {\mathit{DL}}{\xspace}}

\newcommand{\A}{$\mathcal{A}$}
\newcommand{\F}{$\mathcal{F}$}
\newcommand{\server}{\mathcal{S}}

\newcommand{\nab}{\ensuremath {\overline{\mathit{E1}}}{\xspace}}
\newcommand{\forge}{\ensuremath {\mathit{E2}}}{\xspace}
\newcommand{\nabb}{\ensuremath {\mathit{\overline{E3}}}{\xspace}}
\newcommand{\suc}{\ensuremath {\mathit{Win}}{\xspace}}

\newcommand{\hsim}{\ensuremath {\mathit{H}\mhyphen\mathit{Sim}}{\xspace}}

\newcommand\tab[1][1cm]{\hspace*{#1}}
\newcommand{\algrule}[1][.2pt]{\par\vskip.5\baselineskip\hrule height #1\par\vskip.5\baselineskip}
\newcommand{\specialcell}[2][c]{
	\begin{tabular}[#1]{@{}c@{}}#2\end{tabular}}

\usepackage{mathtools, stmaryrd}
\usepackage{xparse} \DeclarePairedDelimiterX{\Iintv}[1]{\llbracket}{\rrbracket}{\iintvargs{#1}}
\NewDocumentCommand{\iintvargs}{>{\SplitArgument{1}{,}}m}
{\iintvargsaux#1} %
\NewDocumentCommand{\iintvargsaux}{mm} {#1\mkern1.5mu,\mkern1.5mu#2}
\newcommand{\romannum}[1]{\uppercase\expandafter{\romannumeral #1\relax}}
\mathchardef\mhyphen="2D 

\newcolumntype{P}[1]{>{\centering\arraybackslash}p{#1}}
\newcolumntype{M}[1]{>{\centering\arraybackslash}m{#1}}

\usepackage{theorem}
\newtheorem{theorem}{Theorem}
{\theorembodyfont{\rmfamily}
	\newtheorem{assumption}{Assumption}{\bfseries}{\rmfamily}}
{\theorembodyfont{\rmfamily}
	\newtheorem{remark}{Remark}{\bfseries}{\rmfamily}}
{\theorembodyfont{\rmfamily}
	\newtheorem{definition}{Definition}{\bfseries}{\rmfamily}}
{\theorembodyfont{\rmfamily}
	{\bfseries}{\rmfamily}}
{\bfseries}{\rmfamily}
{\bfseries}{\rmfamily}
{\bfseries}{\rmfamily}

\newcommand{\mr}[1]{{#1}}

\usepackage[
top    = 0.75in,
bottom = 1in,
left   = 0.625in,
right  = 0.625in]{geometry}

\title{Lightweight and Resilient Signatures for Cloud-Assisted Embedded IoT Systems}

\date{}

\begin{document}
	
	\markboth{}%
	{How to Use the IEEEtran \LaTeX \ Templates}

	\author{Saif E. Nouma \orcidlink{0000-0001-8043-1684}, and Attila A. Yavuz \orcidlink{0000-0002-8680-9307}
		\thanks{This work is supported by Cisco Research Award (220159) and National Science Foundation NSF-SNSF 2444615.  Saif E. Nouma and Attila A. Yavuz are with the Department of Computer Science, University of South Florida, Tampa, 33620, Florida, USA (e-mail: saifeddinenouma@usf.edu, attilaayavuz@usf.edu)}}

\IEEEtitleabstractindextext{%

\begin{abstract}
Digital signatures provide scalable authentication with non-repudiation and are vital tools for the Internet of Things (IoT). Many IoT applications harbor vast quantities of resource-limited devices often used with cloud computing. However, key compromises (e.g., physical, malware) pose a significant threat to IoTs due to increased attack vectors and open operational environments. Forward security and distributed key management are critical breach-resilient countermeasures to mitigate such threats. Yet forward-secure signatures are exorbitantly costly for low-end IoTs, while cloud-assisted approaches suffer from centrality or non-colluding semi-honest servers.  In this work, we create two novel digital signatures called {\em Lightweight and Resilient Signatures with Hardware Assistance} (\lrsha) and its Forward-secure version (\flrsha). They offer a near-optimally efficient signing with small keys and signature sizes. We synergize various design strategies, such as commitment separation to eliminate costly signing operations and hardware-assisted distributed servers to enable breach-resilient verification. Our schemes achieve magnitudes of faster forward-secure signing and compact key/signature sizes without suffering from strong security assumptions (non-colluding, central servers) or a heavy burden on the verifier (extreme storage, computation). We formally prove the security of our schemes and validate their performance with full-fledged open-source implementations on both commodity hardware and 8-bit AVR microcontrollers.

\end{abstract}


}

	\maketitle
	\IEEEdisplaynontitleabstractindextext
	\section{Introduction} \label{sec:Introduction}

The Internet of Things (IoT) is a fast-growing networked system that comprises of vast number of resource-constrained devices (e.g., RFID tags, sensors) \cite{shafique2020internet}. The IoT applications involve domains like health, economy, personal, and military. As a result, the security of IoT devices is critical to achieving trustworthy cyber infrastructures. It becomes even more important when cloud servers are becoming the main resort of the sensitive data collected by IoT devices. The data and its surrounding security services are offloaded to the cloud-enabled ecosystems through emerging data analytic applications.  For example, digital twins aim to conceive virtual replicas of cyber-physical systems (e.g., humans, institutions) \cite{el2021potential} by monitoring the physical entities via IoT equipment (e.g., cameras, sensors, wearable).  Healthcare digital twins deploy various wearables on patients to model and analyze medical functions with IoT devices \cite{marin2016security}. For instance, resource-limited IoT devices (e.g., pacemakers) send electrical pulses to correct a slow heartbeat rate. Additionally, it enables professionals to monitor the patient's health status to prevent heart failures \cite{adamson2009pathophysiology}. Some of these digital twin applications and their security services use cloud assistance \cite{el2021potential}.

Authentication and integrity are vital requirements to guarantee trustworthy  IoT-supported systems \cite{Yavuz:ESCAR:SecuritySafety, iotauthentication2023rao}.  Yet, it is a challenging task to offer these security services efficiently due to resource limitations, scalability issues, and advanced security requirements against system breaches \cite{mudgerikar2021iot}. Below, we outline some of  {\em the highly desirable properties} that an ideal authentication and integrity mechanism must achieve for embedded IoT systems: 

$\bullet$~{\em Lightweight and scalable  signing}: The vast majority of the embedded IoT devices are resource-limited  (e.g., memory, processing, battery) \cite{mudgerikar2021iot}. Hence, the authentication and integrity mechanisms must be lightweight to respect these limitations. Symmetric-key authentication  (e.g., HMAC~\cite{avoine2020symmetric}) is computationally efficient.  However, due to (pairwise) key distribution and management hurdles, they may not be scalable to large-scale and dynamic IoT applications. Moreover, it does not offer public verifiability and non-repudiation, which are crucial features for dispute resolution and audits. Digital signatures offer scalable and public verifiable authentication via public key infrastructures, which makes them ideal for large-scale IoTs. Yet, standard signatures are costly for low-end IoT devices \cite{Yavuz:DSC2023:vision}. The vast majority of signatures require Expensive Operations  (ExpOps) such as modular exponentiation \cite{seo2020efficient}, Elliptic Curve (EC) scalar multiplication~\cite{peng2021efficient} or lattice-operations~\cite{BLISS}, which are shown to be energy and computation intensive for embedded IoTs \cite{ANT:ACSAC:2021, Yavuz:CNS:2019}. 

Lightweight digital signatures \cite{Yavuz:Signature:ULS:TSC, Yavuz:DSC2023:vision, chen2022lfs, peng2021efficient} aim to minimize ExpOps to permit efficient signing. However, this generally comes at the cost of limits on the number of signatures \cite{Yavuz:HASES:ICC2023}, excessively large public keys~\cite{Yavuz:2012:TISSEC:FIBAF}, heavy memory consumption \cite{onlineyao2012}, weakened security \cite{TVHORSInfocom09} or extra assumptions \cite{Yavuz:HASES:ICC2023, Yavuz:CNS:2019}.  The lightweight signing becomes especially challenging when additional security features such as compromise-resiliency and frequent signing  (e.g., as in digital twins) are needed.

$\bullet$~{\em Key compromise-resiliency at IoT device}:   IoT devices are vulnerable to key compromises via malware or physical access (like a smart-watch left in a public place or a medical handheld device left unattended in a hospital)~\cite{MedicalDevice:Survey:2015:Camara2015272}. 
Forward-security mitigates the impact of key compromises via key evolution techniques~\cite{FssAggNew, Yavuz:CORE:CNS:2020} by preventing already issued signatures from being forged after private key exposures. Thus, forward security is a vital security primitive for a wide range of applications. Most notably, secure logging from IoT devices to a remote cloud server, where attackers frequently attempt to alter log artifacts following system compromise events \cite{mitre-clearlogs, liao2024semantic}. 
However, forward-secure signatures \cite{abdalla2019tightness, hulsing2012forward, FWSecItkis_Reyzin_01, FWSig_BellareMiner99} are significantly more expensive than their conventional counterparts. The signing may involve multiple ExpOps with increased key/signature sizes. Even the optimal generic forward-secure transformations incur a logarithmic factor of cost expansion (excluding hidden constants)~\cite{ForwardSecure_MMM_02}. Hence, it is an extremely difficult task to create forward-secure signatures that are lightweight for the signer without putting exorbitant overhead on the verifier \cite{Yavuz:2012:TISSEC:FIBAF}.

$\bullet$~{\em Compact and resilient operations at IoT device}:   (i) The signature/key sizes must be small to respect the memory constraints of embedded devices. (ii) ExpOps require complex arithmetics, which increase the code size and memory footprint. Moreover, they are shown to be more vulnerable to side-channel attacks than simple arithmetic and hash calls~\cite{espitau2017side}. Therefore, it is desirable to limit signing operations to only basic arithmetics and hash calls to avoid these hurdles. (iii) The low-end IoT devices cannot assume a trusted execution environment, and thus the signing logic should not require such special hardware (e.g., unlike \cite{ouyang2021scb}).

$\bullet$~{\em Resiliency at the Cloud-Assistance Services}:  Many lightweight signatures leverage cloud-assistance to attain efficiency and/or advanced security \cite{Yavuz:CNS:2019, Yavuz:HASES:ICC2023, ANT:ACSAC:2021, Yavuz:Patent:Hardware:2023, liu2022lightweight, wang2023secure}. However, the impacts of such cloud assistance must be assessed carefully. (i) The centralized security assistance is prone to a single point of failure, key escrow, and compromise problems. A distributed architecture can mitigate such risks provided that it does not impede the signer's efficiency. (ii) Decentralized signature assistance assumes semi-honest and collusion-risk-free parties, which may not hold in practice. Moreover, the lack of a cheating detection mechanism  (e.g., a party injecting incorrect values) puts the trust at risk. Therefore, it is necessary to provide resiliency not only at the IoT but also at the cloud-assistance side to ensure a higher level of trust and security. 

There is a significant gap in the state-of-the-art to achieve above properties simultaneously.
\mr{
	It is extremely challenging to devise lightweight authentication primitive that achieves minimal computational overhead with a compact memory footprint. Prior works either incur ExpOps \cite{marin2016security} or offload large storage overhead on verifier, and therefore limiting their feasibility and practicality in actual constrained IoT deployments. 
}
Below, we discuss most relevant state-of-art digital signatures to our work and outline the desirable properties of proposed schemes.

\subsection{Related Work} \label{subsec:RelatedWork}
We now summarize the state-of-the-art techniques that are most relevant to our work. Our proposed schemes are lightweight forward-secure digital signatures for embedded IoTs with breach-resilient and decentralized verifier cloud-assistance.  Hence, we select our counterparts through the lenses of these properties. Given that it is not possible to compare our schemes with every digital signature, we first focus on a broad class of seminal signatures.  Later, we capture forward-secure signatures and lightweight constructions relying on special assumptions. Finally, we discuss some signatures with advanced properties that may receive benefit from our schemes or vice versa.

\vspace{2pt}
\noindent \textbf{\em I) Prominent Class of Digital Signatures}: Below, we outline some of the most foundational signatures used in IoTs (and compare our schemes with them in Section \ref{sec:performance_analysis}).

\vspace{2pt}
\indent  $\bullet$  \underline{\em Elliptic Curve (EC)-based Signatures}:  These are currently considered the most suitable class of schemes for resource-limited devices.  ECDSA~\cite{ECDSA} and Ed25519 \cite{Ed25519} are examples of widely experimented schemes on IoT settings \cite{winderickx2022depth}. Despite their merits, they still incur at least one EC scalar multiplication at the signer (e.g., \cite{chen2022lfs, peng2021efficient, li2020permissioned}). It has been shown that even the most efficient EC signatures can be costly for low-end IoT settings (e.g., 8-bit microcontrollers), with a substantial impact on battery life (e.g., \cite{Yavuz:DSC2023:vision,Yavuz:2012:TISSEC:FIBAF}). In our experiments, we re-confirm this fact and then demonstrate that the overhead becomes impractical for low-end devices when advanced features (e.g., forward security) are considered.  We further demonstrate the significant performance difference that lightweight signatures can offer over signatures relying on EC scalar multiplications in signing. 

\vspace{2pt}
\indent \underline{\em $\bullet$ Pairing-based Signatures}: They offer some of the most compact signature and key sizes along with (cross-user) aggregation capability.  Seminal pairing-based schemes like BLS \cite{BLS:2004:Boneh:JournalofCrypto} have been used in various applications such as secure routing \cite{IDBasedAggSig_CCS07}, logging \cite{FssAggNew}, blockchains \cite{drijvers2020pixel}, and IoTs  \cite{liu2022lightweight}. Despite their compactness, the signature generation uses map-to-point and scalar multiplication, which are significantly slower than EC-based schemes. For example, we have shown that BLS signing is  18$\times$ slower than Ed25519, while other studies confirm performance hurdles of BLS on performance-aware networked settings~\cite{baee2021efficiency, gouvea2012efficient}. Hence, we will focus on outperforming EC-based signatures in our work.

\vspace{2pt}
\indent \underline{\em $\bullet$ RSA Signatures}: It achieves a fast verification but highly expensive signing and large key sizes. It is even costlier than BLS-based signatures with larger keys, and therefore is not an ideal choice for our applications~\cite{seo2020efficient, vollala2024energy}.

\vspace{2pt}
\noindent \textbf{\em II) Signatures with Additional Properties and Assumptions}: 

\vspace{2pt}
\indent \underline{\em $\bullet$ Offline/Online (OO) and Pre-Computed Signatures}: These schemes shift expensive signing operations (e.g., EC-scalar multiplication) to an offline phase, thereby permitting faster signing but with extra storage and transmission. The generic OO schemes involve one-time signatures (e.g., HORS \cite{Reyzin2002}) or special hashes (e.g., \cite{OfflineOnline_ImprovedShamir_2001}), which are expensive for low-end devices. Some signatures such as ECDSA \cite{ECDSA} and Schnorr \cite{Schnorr91} naturally permit commitments to be pre-computed \cite{Yavuz:Signature:ULS:TSC}, but require the signer to store a pre-computed token per message (i.e., linear storage overhead). Moreover, after depletion, the signer must re-generate these tokens. Due to these memory/bandwidth hurdles and replenishment costs, such OO approaches  are not suitable for our target applications. 

The BPV techniques \cite{BPV:basepaper:1998} permit a signer to store a pre-computed table, from which commitments can be derived with only EC-scalar additions instead of an EC-scalar multiplication. It has been extensively used in low-end IoTs \cite{Yavuz:CNS:2019, ateniese2013low}. However, recent attacks \cite{coron2020polynomial} on BPV  demands substantially larger security parameters, which reduces the performance gains. There are also pre-computation methods (e.g., \cite{SCRA:Yavuz})  that speed up RSA and BLS, which require a large table storage and scalar additions (for BLS).  A different line of work eliminates the commitment overhead from the signer by relying on a pre-defined set of one-time public keys at the verifier (e..g, \cite{Yavuz:Signature:ULS:TSC, Zaverucha, Yavuz:2012:TISSEC:FIBAF}).  Although signer efficient, they limit the number of signatures to be computed and incur a very large public key storage. 

\vspace{2pt}
\indent \underline{\em $\bullet$ Lightweight Signatures with Cloud Assistance}: Cloud-assistance is used to elevate security in various protocols \cite{Yavuz:Patent:Hardware:2023, liu2022lightweight, wang2023secure, el2021potential, zhang2019efficient}. Various strategies are used to attain lightweight signatures with cloud assistance. In one line, a set of distributed servers supply verifiers with one-time commitments (e.g., \cite{Yavuz:CNS:2019, ANT:ACSAC:2021}). EC-based schemes~\cite{Yavuz:CNS:2019} achieve high computational efficiency but with large key sizes due to BPV. Moreover, the servers are assumed to be semi-honest and non-colluding, without an explicit authentication on the commitment. 
Zhang et al. \cite{zhang2019efficient} propose a certificateless cloud-assisted digital signature scheme, with security based on the Strong Diffie-Hellman (SDH) assumption. While cloud support offloads the computational burden compared to prior works, the scheme still requires at least one scalar multiplication on resource-constrained IoT signers. Therefore, its efficiency is safely reduced to the EC-based class of signatures.

\vspace{2pt}
\noindent \textbf{\em III) Forward-secure (FS) Digital Signatures}:  Seminal FS signatures such as Bellare-Miner \cite{FWSig_BellareMiner99},  Itkis-Reyzin \cite{FWSecItkis_Reyzin_01}, and Abdalla-Reyzin \cite{ForwardSecAbdalla} led to several asymptotically efficient designs (e.g., \cite{FssAggNew, Yavuz:CORE:CNS:2020}). However, they all require (generally multiple) ExpOps at the signing with large signatures, and therefore are not suitable for our use cases.  Another alternative is to transform efficient EC-based signatures into FS with generic transformations (e.g., MMM \cite{ForwardSecure_MMM_02}). MMM is an asymptotically optimal scheme that can transform any signature into an FS variant. However, it requires multiple calls to the underlying signing and key generation, leading to highly costly operations \cite{Yavuz:CORE:CNS:2020}. Akin to MMM, FS signatures such as XMSS \cite{cooper2020recommendation} (and signer-efficient variants  \cite{hulsing2012forward}) also rely on tree structures to attain multiple-time signatures from one-time hash-based schemes. However, as shown in our experiments, they are still magnitudes costlier than our constructions.  Finally, to transform one-time signatures with multiple-time FS schemes, there are FS OO techniques (e.g., \cite{Yavuz:2012:TISSEC:FIBAF}) and cloud-assisted approaches  (e.g., \cite{ANT:ACSAC:2021, Yavuz:HASES:ICC2023}). Despite their signing efficiency, they inherit limited usability, large public keys, and/or risks of single-point failure, as discussed above. Hence, there is a crucial need for FS signatures that avoid ExpOps without strict limits on the number of signatures or heavy one-time public key storage.

\vspace{2pt}
\noindent {\em \textbf{ IV) Lighweight Signatures and Other Cryptographic Constructions with Trusted Execution Environment (TEE)-supported Cloud Assistance}}: 
Numerous digital signature schemes harness TEE-enabled cloud infrastructures to offload the computational overhead inherent in traditional software-only systems. For example, SCB \cite{ouyang2021scb} is a TEE-supported asymmetric cryptographic constructions based on symmetric cryptographic primitives. Although it significantly reduces ExpOps by leveraging lightweight symmetric algorithms, it assumes TEE availability on both endpoints (e.g., signer and verifier in a digital signature setting). Therefore, SCB is infeasible in IoT environments where signers are deemed to be resource-constrained devices (e.g., 8-bit microcontrollers). Another recent digital signature scheme delegates the generation of commitments and public keys to a TEE-supported cloud server~\cite{Yavuz:HASES:ICC2023}. Although it offers lightweight signing and small key sizes, it incurs large signatures. Moreover, this cloud assistance relies on a centralized TEE architecture, therefore prone to the key escrow and central root of trust vulnerabilities (as discussed in \cite{lang2022mole, zhang2019efficient}). 
Conversely, TEEs have been instrumental in numerous cryptographic constructions, beyond digital signatures, most notably Oblivious Random Access Memory (ORAM) in Searchable Encryption (SE) \cite{Yavuz:PETS:2019}, Attribute-Based Encryption (ABE) \cite{arshad2023attribute}, and secure Multi-Party Computation (MPC) protocols \cite{volgushev2019conclave}. These works complement ours and can serve as a privacy-enhancing layer. \vspace{2pt}

\vspace{2pt}
\noindent \textbf{\em V) Alternative Signatures with Potential Extensions}: Identity-based and certificateless signatures \cite{zhou2022efficient} mitigate the overhead of certificate transmission and verification. They have been used in various IoT settings (e.g., \cite{liu2022lightweight, yang2020enhanced, peng2021efficient}). Despite their merits, they still require ExpOp(s) at signers. It is possible to extend our schemes into these settings via proper transformations (e.g., \cite{Yavuz:Certless:ISC:2020}).  

Puncturable digital signatures  (e.g., \cite{jiang2022puncturable, wang2025puncturable}) involve key update strategies and can be built from ID-based signatures (e.g., \cite{jiang2022puncturable}). Multi-signatures (e.g., \cite{chen2022novel}) and threshold signatures (e.g., \cite{sedghighadikolaei2023comprehensive}) can also be extended into forward-secure settings. However, our schemes are signer non-interactive, single-signer, and signer-optimal constructions, and therefore those signatures are not their counterparts. Finally, besides digital signatures, there are myriad other authentication techniques for IoTs, including but not limited to, multi-factor and/or user authentication (e.g., \cite{wang2023secure}). These works are complementary to ours. Our proposed schemes can serve as a building block when used as a signature primitive. Moreover,  our schemes can support a myriad of network services, such as spectrum sensing \cite{grissa2016efficient}. Note that, we strictly aim to guarantee the public verifiability and non-repudiation of the embedded device by itself, but only let the cloud support the verification. Hence, the cloud-assisted authentication methods that defer the signature generation to the cloud (e.g., \cite{zheng2020secure}) are also out of our scope. Finally, the protocols that offer confidentiality and availability for IoTs are out of our scope. 

Given the limitations inherent in digital signature schemes and the research gap in simultaneously achieving the desirable performance and security goals, there is a critical need for a lightweight and resilient digital signature that leverage the existence of cloud servers and advancements in secure hardware (e.g., TEE) present in IoT ecosystems.  
This work attempts to address the following research questions:

\mr{
	\em \textbf{RQ1.} Can we design a digital signature scheme that ensures computational efficiency and minimal energy consumption, enabling practical deployment on IoT endpoints with memory and power constraints?
	
	\textbf{RQ2.} Can such a scheme be realized without incurring substantial storage overhead at the verifier or relying on strong, often impractical, trust assumptions about third-party cloud-assisted servers?
	
	\textbf{RQ3.} Is it possible to leverage TEE support to offload ExpOps during signature generation while avoiding the central root of trust, single-point of failure, and rogue key attacks?
	
	\textbf{RQ4.} Can we achieve FS for signers and breach resilience in cloud entities without affecting performance efficiency?
}


\subsection{Our Contributions}
In this paper, we propose two new digital signatures called {\em Lightweight and Resilient Signatures with Hardware Assistance} (\lrsha) and its forward-secure version as {\em Forward-secure} \lrsha~(\flrsha). Our schemes provide lightweight signing and near-optimally efficient forward security with small keys and signature sizes. They achieve this without relying on strong security assumptions (such as non-colluding or central servers) or imposing heavy overhead on the verifier (like linear public key storage or extreme computation overhead). Our methods introduce and blend different design strategies in a unique manner to achieve these advanced features simultaneously. Some key strategies include using the commitment separation method to eliminate expensive commitment generations and EC operations from the signer and utilizing distributed TEE-supported cloud-assisted servers to provide robust and dependable verification support at the verifier. \mr{We give the details of our schemes in Section \ref{sec:proposed_schemes} and outline the main idea  and design principles of our proposed schemes further below:}

\mr{
	\textbf{\em Main Idea.} 
	The main performance bottleneck of EC-based signature generation arises from the elliptic curve scalar multiplication needed to generate commitments. Prior works (e.g., \cite{Yavuz:2013:EET:2462096.2462108, Yavuz:HASES:ICC2023, Yavuz:CNS:2019}) attempted to mitigate this cost through various commitment management strategies—such as using non-colluding distributed servers \cite{Yavuz:CNS:2019}, a centralized root-of-trust server \cite{Yavuz:HASES:ICC2023}, or accepting linear storage overhead at the verifier side \cite{Yavuz:2013:EET:2462096.2462108}.
	Our proposed scheme, \lrsha, removes the commitment-generation burden from the signer by introducing a distributed commitment strategy built upon a set of TEE-supported servers, called \comc~servers. Each \comc~server securely provides authenticated one-time EC commitments to verifiers on-demand or in batch during an offline phase.
	Furthermore, we extend \lrsha~to a forward-secure variant, \flrsha, which achieves breach resilience against key-compromise attacks through efficient key evolution without relying on heavy certification structures. To the best of our knowledge, (\texttt{F})\lrsha~are the first EC-based signature scheme that simultaneously achieves highly efficient signing, authenticated distributed verification, and collusion resilience.
	We outline the desirable properties of (\texttt{F})\lrsha~further below:
}

\vspace{2pt}
\noindent $\bullet$ \textbf{\underline{\em High Signing Computational Efficiency}}: \lrsha~and \flrsha~provide a near-optimal signature generation with compact key sizes, thanks to the elimination of ExpOps from signing.  \lrsha~outperform their counterparts by being $46\times$ and $4\times$ faster than Ed25519 \cite{Ed25519} and its most signer-efficient counterpart HASES \cite{Yavuz:HASES:ICC2023} on 8-bit AVR ATMega2560 microcontroller. The signing of \flrsha~is also faster than the forward-secure HASES, with a magnitude smaller signature size and without the central root of trust and key escrow limitations. The private key size of \lrsha~is several magnitudes smaller than its fastest counterparts (e.g., \cite{Yavuz:CNS:2019,BPV:Ateniese:Journal:ACMTransEmbeddedSys:2017,SCRA:Yavuz}) that rely on pre-computed tables. \vspace{2pt}

\noindent $\bullet$ \textbf{\underline{\em Forward Security and Tighter Reduction}}:
{\em (i) Forward Security:} As discussed in Section \ref{subsec:RelatedWork}, FS signatures are generally significantly more expensive than their plain variants and not suitable for low-end devices. To the best of our knowledge, \flrsha~is one of the most efficient FS signatures in the literature, whose cost is almost as efficient as few symmetric MAC calls, with a compact signature and key sizes. These properties make it several magnitudes more efficient than existing FS signatures (e.g., \cite{hulsing2012forward,ForwardSecure_MMM_02}) and an ideal  choice to be deployed on embedded IoTs. {\em (ii) Tighter Reduction:} Unlike traditional Schnorr-based signatures (e.g., \cite{Ed25519}), the proof of our schemes avoids the forking lemma, thereby offering a tighter reduction factor. \vspace{2pt}

\noindent $\bullet$ \textbf{\underline{\em Compact, Simple and Resilient Signing}}:  Our signing only relies on a few simple modular additions and multiplication, and cryptographic hash calls. Hence, it does not require intricate and side-channel-prone operations such as EC-scalar multiplication, rejection sampling, and online randomness generation. This simplicity also permits a small code base and memory footprint. These properties increase the energy efficiency of our schemes and make them more resilient against side-channel attacks~(e.g., \cite{genkin2017may, espitau2017side}) targeting complex operations, which are shown to be problematic, especially on low-end IoT devices. \vspace{2pt}

\noindent $\bullet$ \textbf{\underline{\em Collusion-Resilient and Authenticated  Distributed Verif-}} \textbf{\underline{\em -ication with Offline-Online Capabilities:}} Our technique avoids single-point failures and improves the collusion and breach robustness of the verification servers by using a distributed and hardware-assisted signature verification strategy. Furthermore, before signature verification, commitments can be generated and verified offline. In contrast to certain counterpart schemes that depend on servers providing assistance only in a semi-honest, non-colluding, or merely central manner, our systems are able to identify malicious injections of false commitments and provide far quicker signature verification during the online phase. All these characteristics allow our schemes to have reduced end-to-end delays and more reliable authentication than their counterparts with server-aided signatures. \vspace{2pt}


\noindent $\bullet$ \textbf{\underline{\em Full-Fledged Implementation, Comparison, and Validation}}: We implemented our schemes, compared them with our counterparts, and validated their efficiency on both commodity hardware and resource-constrained embedded devices. We open-sourced our full-fledged implementations for reproducibility and future adaptations:
\mr{\fbox{\url{https://github.com/saifnouma/lrsha}}}

\section{Preliminaries}
\label{preliminaries}

The acronyms and notations are described in Table \ref{tab:acronyms}.

\vspace{2pt}

\begin{table}[ht!]
	\caption{List of acronyms and notations }\label{tab:acronyms}
	\centering
	
	\resizebox{0.48\textwidth}{!}{
		\begin{tabular}{|c | l | c | l  | @{}c@{} | @{}c@{} | @{}c@{} | @{}c@{} | @{}c@{} | @{}c@{} | @{}c@{} | @{}c@{} | }
			\hline

			\textbf{Notation/Acronym} &  \textbf{Description}  \\ \hline 
			MCU & Micro-Controller Unit \\ \hline
			TEE & Trusted Execution Environment (also called Secure enclave) \\ \hline
			FS & Forward Security \\ \hline
			(EC)DLP & (Elliptic Curve) Discret Logarithm Problem \\ \hline
			ROM & Random Oracle Model \\ \hline
			EUCMA & Existential Unforgeability against Chosen Message Attack \\ \hline
			(F)HDSGN & FS Hardware-assisted Distributed Signature  \\ \hline
			(F)LRSHA &  FS Lightweight and Resilient Signature with Hardware Assistance  \\ \hline
			PRF & Pseudo-Random Function \\ \hline 
			PPT & Probabilistic Polynomial Time \\ \hline
			\sk/\pk & Private/Public key \\ \hline
			$r$/$R$ & \parbox{0.4\textwidth}{Random nonce/Public commitment (Schnorr-like schemes)} \\ \hline 
			ComC / $C_j$& Commitment Construct and signature $C_j$ on a commitment \\ \hline
			$\server^\ell$ / $a^\ell$ & Identity of the $\ell^{\text{th}}$ ComC server and its private key set \\ \hline
			$\sk'^{\ell}$/$\pk'^{\ell}$ & \parbox{0.4\textwidth}{Private/public keys for certification for  $\ell^{\text{th}}$ ComC server $\server^\ell$} \\ \hline
			$L$ & Number of ComC servers in our system model \\ \hline
			$j / J$ & \parbox{0.4\textwidth}{The algorithm state, and the maximum number of forward-secure signatures to  be generated} \\ \hline 
			$\|$ / $|x|$  &  String concatenation  and bit length of a variable  \\ \hline
			$x \Ra \mathcal{X}$ / $\kappa$ & Random selection from a set $\mathcal{X}$ and security parameter \\ \hline
			$x_j^\ell$ & Variable of server $\server^\ell$ for state  $j$   \\ \hline
			$x_j^{\ell_1,\ell_2}$ & \parbox{0.4\textwidth}{Aggregate variable of  $(x_j^{\ell_1},x_j^{\ell_1+1},\ldots, x_j^{\ell_2})$, where $\ell_2 \ge \ell_1$} \\ \hline
			$\vec{x}$ & \parbox{0.4\textwidth}{Vector contains finite set of elements $\{x_i\}_{i=1}^n$ where $n=|\vec{x}|$ represents of the number of elements in the vector} \\ \hline
			$\{0,1\}^*$ &  Set of binary strings of any finite length  \\ \hline
			${\{q_i\}}_{i=0}^{n}$ & Set of items $q_i$ for $i=0,\ldots, n$   \\ \hline
			$H : \{0,1\}^* \rightarrow  \{0,1\}^{\kappa}$ & Cryptographic hash function \\ \hline
			$H^{(k)}(.)$ & Return the output of $k$ hash evaluations on the same input \\ \hline
		\end{tabular}
	}
\end{table}

\begin{definition} \label{def:sgn}
	A digital signature scheme \sgn~is a tuple of three algorithms $(\kg, \ssig, \ver)$ defined as follows:
	\begin{enumerate}[\indent -]
		\item \underline{$(\sk, \pk, I) \as \sgnkg(1^\kappa)$:} Given the security parameter  $\kappa$, it returns a private/public key pair $(\sk,\pk)$ and system parameters $I$ (implicit input to all other interfaces). 
		\item \underline{$\sigma \as \sgnsig(\sk, M)$:} Given the private key $\sk$ and a message $M$, the signing algorithm returns signature $\sigma$.
		\item \underline{$b \as \sgnver(\pk, M, \sigma)$:} Given the public key $\pk$, message $M$, and a signature $\sigma$, it outputs a bit $b$ (if $b=1$, the signature is valid, otherwise invalid). 
	\end{enumerate}
\end{definition}
 
\begin{definition} \label{def:fs-sgn}
	A forward-secure signature \fsgn~has four algorithms $(\kg, \upd, \ssig, \ver)$ defined as follows:
	\begin{enumerate}[\indent -]
		\item \underline{$(\sk_1, \pk, I) \as \fsgnkg(1^\kappa, J)$:} Given $\kappa$ and the maximum number of key updates $J$, it returns a private/public key pair $(\sk_1,\pk)$ and system parameters $I$ (including state $St\as (j=1)$). 
		\item \underline{$\sk_{j+1} \as \fsgnupd(\sk_j, J)$:} If $j \ge J $ then abort, else, given $\sk_j$, it returns  $\sk_{j+1}$, delete $\sk_j$ and $j \as j+1$. 

		\item \underline{$\sigma_j \as \fsgnsig(\sk_j, M_j)$:}   If $j > J$ then abort, else it computes $\sigma_j$ with $\sk_j$ on $M_j$,  and $\sk_{j+1} \as \fsgnupd(sk_j,J)$.
		
		
		
		\item \underline{$b_j \as \fsgnver(\pk, M_j, \sigma_j)$:} If $j > J$ then abort, else given $\pk$, $M_j$, and $\sigma_j$, it outputs a validation bit $b_j$ (if $b_j=1$, the signature is valid, otherwise invalid). 
	\end{enumerate}
\end{definition}


\begin{definition}
	\label{def:dlp}
	Let $\mathbb{G}$ be a cyclic group of order $q$, $\alpha$ be a generator of $\mathbb{G}$, and \dlp~attacker \A~be an algorithm that returns an integer in $\mathbb{Z}_{q}^{*}$. 
	We consider the following experiment:
	
	Experiment $Expt_{\mathbb{G}, \alpha}^{DL}(\mathcal{A})$:
	\newline 
	\tab $y \Ra \mathbb{Z}_q^*$, $Y \as \alpha^{y} \mod q$,
	$y' \as \mathcal{A}(Y)$,
	\newline 
	\tab If $\alpha^{y'} \mod p = Y$, then return 1, else return 0
	\newline
	The \dl~advantage of \A~in this experiment is defined as:
	\newline
	\tab $\advAdll = Pr[ Expt_{\mathbb{G}, \alpha}^{DL}(\mathcal{A}) = 1]$
	\newline
	The \dl~advantage of $(\mathbb{G}, \alpha)$ in this experiment is as follows:
	\newline
	\tab $\advdll = \underset{\mathcal{A}}{\max}\{\advAdll\}$, where the maximum is over all \A~having time complexity $t$.
\end{definition}

\begin{remark}
	Although we give some definitions for DLP, our implementation is based on Elliptic Curves (EC) for efficiency, and the definitions hold under ECDLP~\cite{costello2016schnorrq}. 
\end{remark}

	\section{System, threat, and security models}
\label{sec:models}

\subsection{System Model}
\label{system-model}

As shown in Figure \ref{fig:systemmodel}, our system model has three entities:

\begin{figure}[ht!]
	\centering
	\includegraphics[width=85mm]{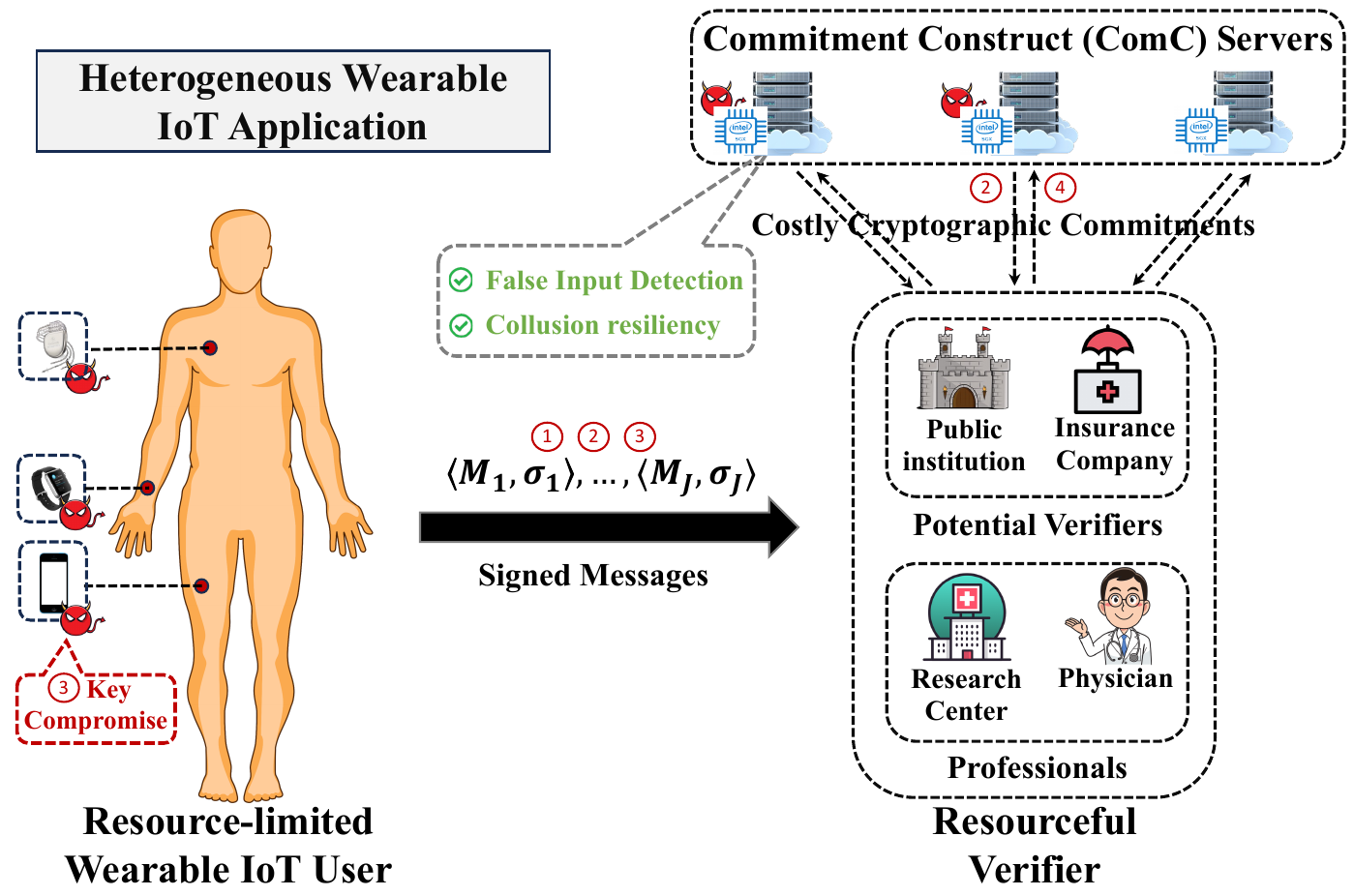}
	\caption[Caption for LOF]{Our System Model}
	\label{fig:systemmodel}
\end{figure}

\begin{enumerate}[1)]
	\item {\em Resource-limited Signer:} We focus on low-end IoT devices as signers.  As depicted in Figure \ref{fig:systemmodel}, we consider a secure wearable medical IoT application, in which the patient is equipped with sensors (e.g., a pacemaker) and wearable devices (e.g., a smart watch) that generate digital signatures on sensitive data to be authenticated by verifiers. 
	

	\item {\em Verifiers:} They can be any entity receiving the message-signature pair from the signer. In our applications, verifiers (e.g., doctors, researchers, insurance companies) are equipped with commodity hardware (e.g., a laptop).  
	

	\item {\em \comc~Servers:} \comc~servers $\server= (\server^1, \ldots, \server^L)$, where each server is equipped with a TEE (i.e., secure enclave). We used Intel SGX due to its wide availability (e.g., Microsoft Azure). However, our model can be implemented with any TEE  (e.g., ARM TrustZone, Sanctum).
	
	
\end{enumerate}

\subsection{Threat and Security Model} \label{security-model}
\label{threat-model}
Our threat model is based on an adversary with the following capabilities:
\begin{enumerate}[1)]
	\item {\em Passive attacks}: Monitor and interpret the output of the cryptographic interfaces sent from the IoT end devices and/or the \comc~servers. 
	\item {\em Active attacks}: Attempt to intercept, forge, and modify messages, signatures and auxiliary values (e.g., commitments) sent from IoT devices and \comc~servers.
	\item {\em Key Compromise  - resource-limited IoT side}: Attempt breaching device to extract the cryptographic secret~\cite{mudgerikar2021iot}.
	\item {\em Breach attempts on assisting clouds for verification services}: 	Attempt to gain access to assisted cloud services to tamper with the protocol such that:
	(i) Inject incorrect commitments.
	(ii) Forge certificates of the commitments.
	(iii) Force the cloud to collude (e.g., expose the secret keys).
\end{enumerate}
Below, we first define the interfaces of our proposed schemes, and then present their security model that captures the above threat model as follows:

\begin{definition}
	\label{def.hsgn}
	A hardware-assisted distributed digital signature scheme \hdsgn~consists of four algorithms ($\kg, \comconstr, \sgn, \ver$) defined as follows:

	\begin{enumerate}[\indent -]
		\item \underline{$(\sk, \pk, \vec{a}, I) \as \hdsgnkg(1^\kappa,L)$:} Given $\kappa$ and the number of \comc~servers $L$, it returns a private/public key pair $(\sk,\pk)$, system parameter $(I,St \as j=1)$, and private key of each \comc~server $\vec{a}=\{a^\ell\}_{\ell=1}^L$.
		\item \underline{$ (\vec{R_j}, \vec{C_j}) \as \hdsgncomconstr(\{a^\ell\}_{\ell=1}^{L}, j)$:} Given $St=j$ and $\vec{a}$, each server $\server^\ell$ generates a commitment $R_j^\ell$ and its signature   $C_j^\ell=\sgnsig_{a^\ell}(R_j^\ell)$. \hdsgncomconstr~returns  $(\vec{R_j}=\{R_j^\ell\}_{\ell=1}^L,\vec{C_j}=\{C_j^\ell\}_{\ell=1}^L)$ as output. 
		\item \underline{$\sigma_j \as \hdsgnsig(\sk, M_j)$:} Given $\sk$ and a message $M_j$, it returns a signature $\sigma_j$ and $j\as j+1$.
		\item \underline{$b_j \as \hdsgnver(\pk, M_j, \sigma_j)$:} Given $\pk$, $M_j$, and its signature $\sigma_j$, the verification algorithm calls~$(\vec{R_j}, \vec{C_j}) \as \hdsgncomconstr(\{a^\ell\}_{\ell=1}^{L}, j)$, and then it outputs a bit $b_j$ (if $b_j=1$, the signature is valid, otherwise invalid). 
	\end{enumerate}
\end{definition}

\begin{definition}
	\label{def.hfsgn}
	A forward-secure and hardware-assisted distributed digital signature scheme \fhdsgn~consists of five algorithms ($\kg, \comconstr, \upd, \ssig, \ver$) defined as follows:
	
	\begin{enumerate}[-]
		\item \underline{$(\sk_1, \pk, \vec{a}, I) \as \fhdsgnkg(1^\kappa,J,L)$:} Given $\kappa, L$, and the maximum number of signatures $J$ to be produced, it returns $(\sk_1,\pk)$, $(I, St \as j =1)$, and  $\vec{a}=\{a^\ell\}_{\ell=1}^L$.
		
		\item \underline{$ (\vec{Y_j},\vec{R_j},\vec{C_j}) \as \fhdsgncomconstr(\vec{a}, j)$:} Given $\vec{a}$ and state  $j$, it returns a set of public key and commitment set $(\vec{Y_j}=\{Y_j^\ell\}_{\ell=1}^L, \vec{R_j}=\{R_j^\ell\}_{\ell=1}^L)$, a forward-secure signature on  each pair as $\vec{C_j}=\{\fsgnsig_{a^\ell}(Y_j^\ell \| R_j^\ell)\}_{j=1}^{L}$, and returns $(\vec{R_j},\vec{C_j})$.  
		
			\item \underline{$\sk_{j+1} \as \fhdsgnupd(\sk_j, J)$:} As in Definition~\ref{def:fs-sgn} update. 
			
			
	\item \underline{$\sigma_j \as \fhdsgnsig(\sk_j, M_j)$:} As in Definition \ref{def:fs-sgn} signing. 
	

	\item \underline{$b_j \as \fhdsgnver(\pk, M_j, \sigma_j)$:} If $j > J$ then abort. Otherwise, given the public key $\pk$, a message $M_j$, and its signature $\sigma_j$, the verification algorithm calls~$(\vec{Y_j}, \vec{R_j}, \vec{C_j}) \as  \fhdsgncomconstr(\vec{a}, j)$, and then it outputs a bit $b_j$ (if $b_j=1$, the signature is valid, otherwise invalid). 
	\end{enumerate}
\end{definition}

The standard security notion for a digital signature \sgn~is the Existential Unforgeability against Chosen Message Attack (\EUCMA)~\cite{Yavuz:CNS:2019}. It captures a Probabilistic Polynomial Time (PPT) adversary \A~aiming at forging signed messages. It corresponds to capabilities (1-2) stated in the threat model (passive or active attacks on message-signature pairs). 

\begin{definition}\label{def:EUCMA}
	\EUCMA~experiment $ \mathit{Expt}^{\EUCMA}_{\sgn} $ for  $  \sgn$ is as follows (in random oracle model (ROM)~\cite{katz2020introduction}):
	\begin{enumerate}[\indent -]
		\item  $ (\sk,\pk,I)\leftarrow \sgnkg(1^\kappa) $
		\item $ (M^*, \sigma^*)\leftarrow \mathcal{A}^{\ro, ~\Sgnsig } (\pk) $
	\end{enumerate}
		
	\A~wins the experiment if  $ {\sgnver(\pk, M^*, \sigma^*)} = 1$ and $ M^* $ was not queried to  $\Sgnsig$ oracle.
	The \EUCMA~advantage of \A~is defined as $\advsgnA = \Pr[ \mathit{Expt}^{\EUCMA}_{\sgn}= 1]$. The \EUCMA~advantage of \sgn~is defined as $\advsgn=\max_{\mathcal{A}}~\advsgnA$. Note that the maximum is evaluated across all of the possible \A~with time complexity $t$ and maximum number of running queries $q_H$ and $q_s$ to the \ro~and \Sgnsig~oracles, respectively. 
		\begin{enumerate}
		\item {\em Random Oracle \ro}: It handles \A's hash queries on any message $M$ by returning a randomly uniformly distributed output $h \as \mathit{RO}(M)$. All cryptographic hashes used in our schemes are modeled as \ro~\cite{katz2020introduction}. 
		\item {\em $\sgnsig_{\sk}(.)$ }: It provides a signature $\sigma$ on any queried message $M$ computed as  $\sigma \as \sgnsig_{\sk}(M)$. 
	\end{enumerate}	
\end{definition}


We follow the formal security model of a hardware-assisted distributed digital signature scheme (\hdsgn) as the Hardware-assisted Distributed Existential Unforgeability against Chosen Message Attack (\HDEUCMA). It captures the capabilities (1-2, 4) in our threat model, including  \A's potential attacks on the \comc~servers. 


\begin{definition}\label{def:HDEUCMA}
	\HDEUCMA~experiment $ \mathit{Expt}^{\HDEUCMA}_{\hdsgn} $ for a hardware-assisted distributed digital signature $  \hdsgn = (\kg, \comconstr, \ssig, \ver) $ is defined as follows:
	\begin{enumerate}[\indent -]
		\item  $ (\sk,\pk,\{a^\ell\}_{\ell=1}^L,I)\leftarrow \hdsgnkg(1^\kappa,L) $
		\vspace{1pt}
		\item \resizebox{.89\hsize}{!}{$ (M^*, \sigma^*)\leftarrow \mathcal{A}^{ \ro,~\Hdsgnsig,~\Hdsgncomconstr } (\pk) $} 
	\end{enumerate}
\noindent, where $\vec{a}=\{a^\ell\}_{\ell=1}^{L},$ denotes private key material of ComC server $\{\server^\ell\}_{\ell=1}^{L}$, respectively. 


	\vspace{2pt}
	
	\A~wins the experiment if  $ {\hdsgnver(\pk, M^*, \sigma^*)} = 1$ and $ M^* $ was not queried to  $\Hdsgnsig$. 
	The \HDEUCMA~advantage of \A~is defined as $\advhdsgnA = \Pr[ \mathit{Expt}^{\HDEUCMA}_{\hdsgn}= 1]$.  The \HDEUCMA~advantage of \hdsgn~is defined as $\advhdsgn=\max_{\mathcal{A}} \advhdsgnA$ with all possible adversary \A~having time complexity $t$ and maximum queries $q_H$ to \ro~and $q_s$ to both of \Hdsgnsig~and \Hdsgncomconstr.  
			\begin{enumerate}
		\item Oracles \ro~and \Hdsgnsig~works in as Def. \ref{def:EUCMA}. 
		\item {\em  \comconstral}: Given state $j$, it generates a public commitment $\{R_j^\ell\}_{\ell=1}^L$ and corresponding signature $\{C_j^\ell \as \sgnsig_{a^{\ell}}(R_j^\ell) \}_{\ell=1}^L$ for each ComC server $\{\server^\ell\}_{\ell=1}^L$.
		
	\end{enumerate}	
	
\end{definition}


The standard security notion for a forward-secure digital signature scheme \fsgn~is the Forward-secure \EUCMA~(\FEUCMA) \cite{FWSig_BellareMiner99}. It captures the key compromise capability 3) in our threat model.  

\begin{definition}\label{def:FEUCMA}
	\FEUCMA~experiment $ \mathit{Expt}^{\FEUCMA}_{\fsgn} $ for a forward-secure signature scheme $ \fsgn = (\kg,\upd,\ssig, \ver)$ is defined as follows:
	\begin{enumerate}[\indent -]
		\item  $ (\sk_1,\pk,I) \leftarrow \fsgnkg(1^\kappa,J) $
		\item $(M^*, \sigma^*)\leftarrow \mathcal{A}^{\ro,~\Fsgnsigj,~\Breakin} (\pk) $
	\end{enumerate}
	
	\A~wins the experiment if  $ {\fsgnver(\pk, M^*, \sigma^*)} = 1 $ and $ M^* $ was not queried to $\Fsgnsigj$. The \FEUCMA~advantage of  \A~is defined as $\advfsgnA = \Pr[ \mathit{Expt}^{\textit{\FEUCMA}}_{\fsgn}= 1]$. The \FEUCMA~advantage of \fsgn~is defined as $\advfsgn = \max_{\mathcal{A}} \advfsgnA$, with all possible \A~having time complexity $t$ and $q_H$, $q_s$, and one queries to \ro, \Fsgnsigj, and \Breakin~oracles, respectively. \ro~and \Fsgnsigj~oracles are as in Definition \ref{def:EUCMA}. \Breakin~oracle returns the private key $\sk_{j+1}$ if queried on state $1 \le j < J$, else aborts. 
\end{definition}


We follow the formal security model of a forward-secure hardware-assisted distributed digital signature scheme (\fhdsgn) as Forward-secure \HDEUCMA~(\FHDEUCMA).  It combines both security definitions and captures all abilities of the attacker (1-4) in our threat model.  {\em This offers improved security over non-forward secure signature and/or cloud-assisted signature schemes that only rely on a semi-honest model}.


\begin{definition}\label{def:FHDEUCMA}
	\FHDEUCMA~experiment $ \mathit{Expt}^{\FHDEUCMA}_{\fhdsgn} $ for a forward-secure and hardware-assisted signature $ \fhdsgn = (\kg,\comconstr, \upd,\ssig,\ver)$ is defined as follows:
	\begin{enumerate}[\indent -] 
		\item  $ (\sk_1,\pk,I) \leftarrow \fhdsgnkg(1^\kappa,J,L) $
		\item \resizebox{.89\hsize}{!}{ $(M^*, \sigma^*)\leftarrow \mathcal{A}^{\ro,~\Fhdsgnsig,~\Fhdsgncomconstr,~\Breakin} (\pk) $ }
	\end{enumerate} 
	
	\vspace{4pt}
	
	\A~wins the experiment if  $ {\fhdsgnver(\pk, M^*, \sigma^*)} = 1 $ and $ M^* $ was not queried to $\Fhdsgnsig$ oracle. The \FHDEUCMA~advantage of \A~is defined as $\mathit{Adv}^{\FHDEUCMA}_{\fhdsgn} = \Pr[ \mathit{Expt}^{\textit{\FHDEUCMA}}_{\fhdsgn}= 1]$.
	The \FHDEUCMA~advantage of \fhdsgn~is defined as $\advfhdsgn = \max_{\mathcal{A}}~\advfhdsgnA$, with all possible \A~having time complexity $t$ and maximum queries equal to $q_H$, $q_s$, and one to \ro, both of \Fhdsgnsig~and \Fhdsgncomconstr, and \Breakin, respectively.  All oracles behave as in Def. \ref{def:HDEUCMA}, except FS signatures are used for signing in \Fhdsgnsig, \Fhdsgncomconstr~and \Breakin~(as in Definition \ref{def:FEUCMA}).
\end{definition}

\begin{assumption} \label{assump:tee2}
	Each \comc~server $\{\server^\ell\}_{\ell=1}^{L}$ securely provisions secret keys (before deployment) and runs their commitment construction functions via a secure Trusted Execution Environment (TEE) as described in the system model to offer colluding resistance and commitment authentication. 
\end{assumption}

	{\em Discussion}: The malicious security properties (i.e., capability 4 in our threat model) of  our schemes at the assisting servers rely on the security of the underlying TEE. This offers enhanced mitigation to collusion and malicious tampering attacks on the stored private keys on the \comc~servers. This is realized with a low cost and without having any impact on the signer performance. Unlike some related works (e.g., \cite{ouyang2021scb}), we do not require a TEE on the signer.  We realized our TEE with Intel SGX's secure enclaves. However, our system could also be instantiated using other isolated execution environments (e.g., Sanctum \cite{costan2016sanctum}). It is crucial to recognize the limitations of relying on trusted execution environments. For example, Intel SGX encountered various side-channel attacks (e.g., \cite{silva2022power}). Generic techniques for protection against enclave side-channel attacks are also under study in various works (e.g., \cite{lang2022mole}), therefore they are complementary to ours. Finally,  even if TEE on some \comc~servers is breached, the EU-CMA property of our \lrsha~scheme will remain as secure as our counterpart cloud-assisted signatures (e.g., ~\cite{ANT:ACSAC:2021, Yavuz:CNS:2019}), which assume a semi-honest server model with  $(L-1,L)$-privacy.  However, our forward-secure scheme in this case can only achieve EU-CMA as \lrsha. We further note that \A~successfully launching side-channel attacks against multiple TEEs on distinct ComC servers simultaneously assumes an extremely strong adversary.

\section{Proposed Schemes} \label{sec:proposed_schemes}
We first outline our design principles and how we address some critical challenges of constructing a highly lightweight signature with hardware-supported cloud assistance. We then describe our proposed schemes in detail. 




\textbf{High-Level Idea and Design Principles}: Fiat-Shamir type EC-based signatures (e.g., Ed25519~\cite{Ed25519}, FourQ~ \cite{costello2016schnorrq}) are among the most efficient and compact digital signatures. Their main overhead is the generation of a commitment $R\as \alpha^r \mod p$ (EC scalar multiplication) from one-time randomness $r$. In Section \ref{subsec:RelatedWork}, we captured the state-of-the-art lightweight signatures that aim to mitigate this overhead via various commitment management strategies.  

In our design, we exploit the commitment separation method~(e.g., \cite{Yavuz:2013:EET:2462096.2462108, nouma2023practical}), but with various advancements to address the challenges of previous approaches.  In commitment separation, the value $R_j$ in $H(M_j\|R_j)$  is replaced with one-time randomness $x_j$ per message as $H(M_j\|x_j)$.  This permits $R_j$ to be stored at the verifier before signature generation, provided that $x_j$ is disclosed only after signing. While this approach eliminates ExpOps due to commitment generation, it has significant limitations: (i) The verifier must store a commitment per message to be signed that incurs linear public key storage overhead  \cite{Yavuz:Signature:ULS:TSC,Yavuz:2012:TISSEC:FIBAF} (i.e., one-time commitments become a part of the public key). (ii) This limits the total number of signatures to be computed and puts a burden on signers to replenish commitments when depleted. 

Our strategy is to {\em completely} eliminate the burden of commitments from the signer, but do so and by achieving advanced security properties such as {\em forward-security} and {\em malicious server detection with collusion-resistance}, which are not available in previous counterparts simultaneously:

 (i) Our design uses a distributed commitment strategy, in which value $r$ is split into $L$ different shares $\{r^{\ell}\}_{\ell=1}^{L}$ each provided to a TEE-supported \comc~server $\{\server^{\ell}\}_{\ell=1}^{L}$ along with other keys to enable advanced features (to be detailed in algorithmic descriptions). This approach mitigates single-point failures and key compromise/escrow problems in centralized cloud-assisted designs  (e.g.,~\cite{Yavuz:HASES:ICC2023,Yavuz:Signature:ULS:TSC}).  (ii) Our design does not rely on BPV~\cite{ateniese2013low} (unlike~\cite{Yavuz:CNS:2019}) or signature-tables (unlike \cite{SCRA:Yavuz}), but only uses simple arithmetic and PRF operations.  This permits both computational and memory efficiency. If we accept equal table storage as our counterparts, then this further boosts our speed advantage. (iii) Some previous EC-based server-assisted signatures rely on semi-honest servers, which are prone to collusion and lack the ability to detect servers supplying false commitments. Instead,  we wrap our \comc~servers with a TEE that not only mitigates the collusion risk, but also forces the attacker to breach multiple TEE instances to extract keys or coerce an algorithmic deviation.  This substantially increases the practical feasibility of active attacks targeting \comc~servers. Moreover, our \comc~servers authenticate each commitment separately, permitting verifiers to detect the server(s) injecting a false commitment.  With TEE support, after detection, we can also use attestation to further mitigate post-compromise damages. (iv) We have a new forward-secure variant with an efficient key evolution strategy that avoids heavy nested certification trees (e.g., unlike \cite{cooper2020recommendation,ForwardSecure_MMM_02}) and costly public key evolutions (e.g., \cite{FssAggNew}). Thanks to this, our scheme offers more than 15 times faster signing with 24 times smaller signatures compared to the most efficient (generic) forward-secure EC-based counterpart (see Section~\ref{sec:performance_analysis}).
 

%
We now present our schemes \lrsha~and  \flrsha.

\subsection{Lightweight and Resilient Signature with Hardware Assistance (LRSHA)}
\label{subsec:lrsha}

\begin{figure}[ht!]
	\centering
	\includegraphics[width=90mm]{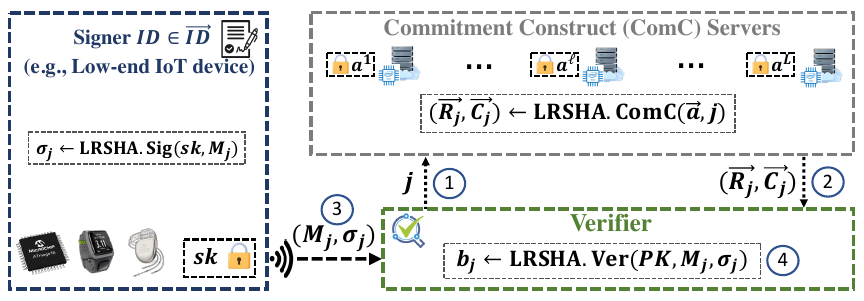}
	\caption[Caption for LOF]{High-Level Overview of \lrsha.}
	\label{fig:lrsha}
\end{figure}

We  created Lightweight and Resilient Signature with Hardware Assistance (\lrsha), which is outlined in Fig. \ref{fig:lrsha}  and detailed in  Alg. \ref{alg:lrsha}.  We further elaborate steps in Alg. \ref{alg:lrsha} as follows.

\begin{algorithm}[ht!]
	
	\scriptsize
	\caption{ Lightweight and Resilient Signature with Hardware Assistance (\lrsha) }
	\label{alg:lrsha}
	
	\begin{algorithmic}[1]
		\Statex   $\underline{(\sk, \pk, \vec{a}, I)\as \lrshakg(1^{\kappa},L)}$:
		\vspace{1pt}
		
		\State Generate large primes $q$ and $p$ such that $q|(p-1)$. Select a generator $\alpha$ of the subgroup $\mathbb{G}$ of order $q$ in $\mathbb{Z}_{q}^{*}$. Set $I \as (p,q,\alpha,St \as j=1)$
		\State $y \Rq$ and $Y \as \alpha^y \mod p$
		
		\For{$\ell=1,\ldots,L$}
		\State $(\sk'^\ell,\pk'^\ell) \as \sgnkg(1^\kappa)$
		\State $r^\ell \Rq$
		\State $a^\ell \as \langle\sk'^\ell, r^\ell \rangle $ is securely provisioned to the enclave of server $\server^\ell$
		\EndFor
		
		\State $\sk = \langle y,\vec{r}=\{r^\ell\}_{\ell=1}^L \rangle $ 
		\vspace{1pt}
		\State \Return ($\sk, \pk=(Y, \vec{\pk'}=\{\pk'^\ell\}_{\ell=1}^L), \vec{a}=\{ a^\ell \}_{\ell=1}^{L}, I$)
	\end{algorithmic}
	\algrule
	\begin{algorithmic}[1]
		\Statex $\underline{ (\vec{R_j},\vec{C_j}) \as \lrshacomconstr( \vec{a}, j )}$: 
		
		\For{$\ell=1,\ldots,L$}
		\State $R_j^\ell \as \alpha^{r_j^\ell} \mod p $ ~, where $r_j^\ell \as \prf_{r^\ell}(j) \mod q$
		\State $C_j^\ell \as \sgnsig(\sk'^\ell, R_j^\ell)$
		\EndFor
		
		\State \Return $( \vec{R} = \{R_j^\ell\}_{\ell=1}^L , \vec{C}=\{C_j^\ell\}_{\ell=1}^L)$
	\end{algorithmic}
	\algrule
	\begin{algorithmic}[1]
		\Statex $\underline{\sigma_j \as \lrshasig(\sk, M_j)}$: 
		\vspace{1pt}
		
		\State $r_j^{1,L} \as \sum_{\ell=1}^{L} r_j^\ell \mod q$~, where $r_j^\ell \as \prf_{r^\ell}(j) \mod q$
		\State $e_j \as H(M_j \| x_j) \mod q$~, where $x_j \as \prf_{y}(j) \mod q$
		\State $s_j \as r_j^{1,L} - e_j \cdot y \mod q$
		\State Update $St \as j+1$
		
		\State	\Return $\sigma_j \as \langle s_j, x_j, j \rangle$
	\end{algorithmic}
	\algrule
	\begin{algorithmic}[1]
		\Statex $\underline{b_j \as \lrshaver( \pk, M_j,\sigma_j)}$:  Step 1-4 can be run offline. 
		\vspace{3pt}
		
		\State $(\vec{R_j}, \vec{C_j}) \as \lrshacomconstr( \vec{a},j)$  \Comment{Offline}
		
		\For{$\ell=1,\ldots, L$} \Comment{Offline}
		\State \textbf{if} $\sgnver(\pk'^\ell, R_j^\ell, C_j^\ell)=1$ \textbf{then continue else return} $b_j=0$
		\EndFor
		\State $R_j^{1,L} \as \prod_{\ell=1}^L {R_j^\ell} \mod p$ \Comment{Offline}
		\State $e_j \as H(M_j \| x_j) \mod q$
		\State \textbf{if} $R_j^{1,L}={\alpha}^{s_j} \cdot Y^{e_j} \mod p$, \textbf{return} $b_j=1$, \textbf{else return} $b_j=0$
	\end{algorithmic}
	
	\vspace{-3pt}
\end{algorithm}

The key generation algorithm \lrshakg~accepts the security parameter $\kappa$  and the number of \comc~servers $L$. It first  generates EC-related parameters $I$ and the main private/public key pair (Step 1-2), and then  a commitment certification private/public key pair ($\sk'^\ell, \pk'^\ell$) for each server $\server^\ell$ (Step 4). Subsequently, it generates private key components $a^\ell = \langle \sk'^\ell,r^\ell \rangle$ to be provisioned to each secure enclave of the server $\server^\ell$ (step 5-6). Finally, \sk~and the internal state $St=(j\as1)$ are provided to the signer (Step 7-8).


In the signature generation algorithm \lrshasig, given the state $j$, the signer first computes $r_j^{1,L}$ by aggregating values $\{r_j^\ell\}_{\ell=1}^L$  via PRF calls (Step 1).  The one-time randomness $x_j$ is used as the commitment (Step 2) instead of the public commitment $R$. Step 3 is as in Schnorr's signature, followed by a state update.  Overall, our signing avoids any ExpOp, costly pre-computed tables (e.g., BPV or signature tables), or secure hardware requirements. 


\lrshaver~is a cloud-assisted  verification algorithm, and therefore calls  \lrshacomconstr~to retrieve  $L$ partial commitment values $\{R_j^\ell\}_{\ell=1}^L$ and their certificates from \comc~servers (Step 1).  In \lrshacomconstr, each server $\server^\ell$  first derives  $\{R_j^\ell\}_{\ell=1}^L$ from their private keys $\vec{a}=\{ a^\ell \}_{\ell=1}^{L}$ (step 2), puts a signature to certify them as  $\{C_j^\ell\}_{\ell=1}^L$ (step 3) and returns these values to the verifier. The rest of \lrshaver~is similar to EC-Schnorr but with randomness $x_j$ instead of commitment $R_j$  in hash (steps 5-6). Note that the verifier can retrieve commitments and verify certificates offline (and even in batch) before message verification occurs. Hence, the overall online message verification overhead is identical to the EC-Schnorr signature. Moreover, \lrshaver~does not require any pre-computed table, lets the verifier detect false commitments, and offers distributed security for assisting servers with enhanced collusion resiliency via TEE support in \lrshacomconstr.

\subsection{Forward-secure Lightweight and Resilient Signature with Hardware Assistance (FLRSHA)} \label{subsec:flrsha}
\begin{figure}[ht!]
	\centering
	\includegraphics[width=90mm]{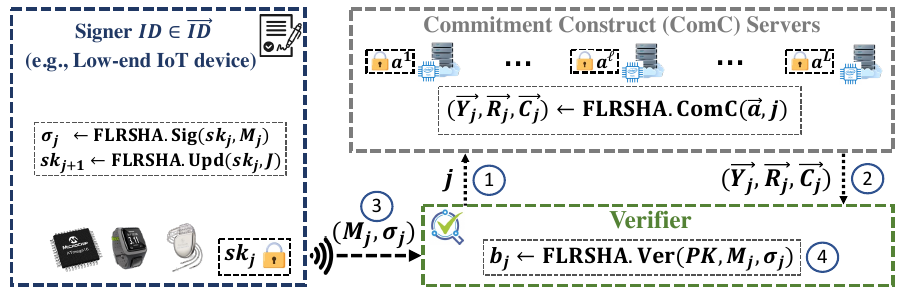}
	\caption[Caption for LOF]{The high-level overview of \flrsha.}
	\label{fig:flrsha}
\end{figure}

We now present our Forward-secure \lrsha~(\flrsha) as detailed in Algorithm \ref{alg:flrsha} with an overview in Figure \ref{fig:flrsha}. We developed a key evolution mechanism for the signer and \comc~servers that enables a highly lightweight yet compromise-resilient digital signature. Our introduction of distributed TEEs provided significant performance and security benefits, making \flrsha~the most efficient forward-secure alternative for low-end embedded devices (see in  Section \ref{sec:performance_analysis}). Below, we outline \flrsha~algorithms by focusing on their differences with \lrsha.



\begin{algorithm}[ht!]
	\scriptsize
	\caption{ Forward-secure Lightweight and Resilient Signature with Hardware Assistance (\flrsha)}
	\label{alg:flrsha}
	
	\begin{algorithmic}[1]
		\Statex   $\underline{(\sk_1, \pk, \vec{a}, I)\as \flrshakg(1^{\kappa},J,L)}$:
		\vspace{1pt}
		
		\State Generate primes $q$ and $p > q$ such that $q | (p - 1)$. Select a generator $\alpha$~of the subgroup $\mathit{G}$~of order $q$~in $\mathbb{Z}_{p}^{*}$. Set $I \as \langle p, q, \alpha \rangle $ as system parameter.
		
		\For{$\ell=1,\ldots,L$}
		\State $(\sk'^\ell, \pk'^\ell) \as \fsgnkg(1^\kappa)$
		\State $y_1^\ell \Rq$ ~, and $r_1^\ell \Rq$
		\State $a^\ell \as \langle y_1^\ell, r_1^\ell, \sk'^\ell \rangle $ is securely provisioned to the TEE of server $\server^\ell$
		\EndFor
		
		\State $\sk_1 \as \langle \vec{y_1}=\{y_1^\ell\}_{\ell=1}^L,\vec{r_1}=\{r_1^\ell\}_{\ell=1}^L \rangle $. The signer's initial state is $St\as j = 1$
		
		\State \Return ($\sk_1, \pk=\{\pk'^\ell\}_{\ell=1}^L~, \vec{a}=\{ a^\ell \}_{\ell=1}^{L}~, I$)
	\end{algorithmic}
	\algrule
	\begin{algorithmic}[1]
		\Statex $\underline{ (\vec{Y_j}, \vec{R_j}, \vec{C_j} ) \as \flrshacomconstr( \vec{a}, j)}$: 
		Each  $S^1,\ldots,S^L$ executes in their independent TEE in isolation. It can be done in batch offline, or on demand online.

		\For{$\ell=1,\ldots,L$}
		\State $Y_j^\ell \as \alpha^{y_j^\ell} \mod p$ ~, where $y_j^\ell \as H^{(j-1)}(y_1^\ell) \mod q$
		\State $R_j^\ell \as \alpha^{r_j^\ell} \mod p $ ~, where $r_j^\ell \as H^{(j-1)}(r_1^\ell) \mod q$
		\State $C_j^\ell \as \fsgnsig(\sk'^\ell, Y_j^\ell \| R_j^\ell)$
		\EndFor
		
		\State \Return $ ( \vec{Y_j}=\{Y_j^\ell\}_{\ell=1}^L, \vec{R_j}=\{R_j^\ell\}_{\ell=1}^L, \vec{C_j}=\{C_j^\ell\}_{\ell=1}^L)$
	\end{algorithmic}
	\algrule
	\begin{algorithmic}[1]
		\Statex $\underline{ \sk_{j+1} \as \flrshaupd(\sk_j, J)}$: If $j \ge J$ then {\em abort} else continue: 
		
		\For{$\ell=1,\ldots,L$}
		\State $y_{j+1}^{\ell} \as H(y_j^\ell) \mod q$
		\State $r_{j+1}^\ell \as H(r_j^\ell) \mod q$
		\EndFor
		\State Set $\vec{y}_{j+1}=\{y_{j+1}^\ell\}_{\ell=1}^L$~, $\vec{r}_{j+1}=\{r_{j+1}^\ell\}_{\ell=1}^L$, and $St \as  j+1$
		\State \Return $\sk_{j+1} \as \langle \vec{y}_{j+1}~, \vec{r}_{j+1} \rangle$
	\end{algorithmic}
	\algrule
	\begin{algorithmic}[1]
		\Statex $\underline{\sigma_j \as \flrshasig(\sk_j, M_j)}$: 	If  $ j>J$ then {\em abort}, else continue:
		\vspace{1pt}
		
		\State $y_j^{1,L} \as \sum_{\ell=1}^L {y_j^\ell} \mod q$ , and $r_j^{1,L} \as \sum_{\ell=1}^{L} r_j^\ell \mod q$
		\State $e_j \as H(M_j \| x_j) \mod q$~, where $x_j \as \prf_{y_j^{1,L}}(j)$
		\State $s_j \as r_j^{1,L} - e_j \cdot y_j^{1,L} \mod q$
		\State $\sk_{j+1} \as \flrshaupd(\sk_j,J)$
		
		\State	\Return $\sigma_j \as \langle s_j, x_j, j \rangle$
	\end{algorithmic}
	\algrule
	\begin{algorithmic}[1]
		\Statex $\underline{b_j \as \flrshaver( \pk, M_j,\sigma_j)}$:  If  $ j>J$ then {\em abort}, else continue: Note that steps 1-4 can be run offline. 
		\vspace{3pt}
		
		\State $ (\vec{Y_j}, \vec{R_j}, \vec{C_j}) \as \flrshacomconstr( \vec{a}, j)$ \Comment{Offline}
		
		\For{$j=1,\ldots, L$}  \Comment{Offline}
		\State \textbf{if} $\fsgnver(\pk'^\ell, Y_j^\ell \| R_j^\ell, C_j^\ell)=1$ \textbf{then continue else return} $b_j=0$
		\EndFor
		\State $Y_j^{1,L} \as \prod_{\ell=1}^L {Y_j^\ell} \mod p$~, and $R_j^{1,L} \as \prod_{\ell=1}^L {R_j^\ell} \mod p$  \Comment{Offline}
		\State $e_j \as H(M_j \| x_j) \mod q$
		\State \textbf{if} $R_j^{1,L}= \alpha^{s_j} \cdot {(Y_j^{1,L})}^{e_j} \mod p$ \textbf{then return} $b_j=1$ \textbf{else return} $b_j=0$
	\end{algorithmic}
	
	\vspace{-3pt}
\end{algorithm}

The key generation \flrshakg~works as in~\lrshakg~but with the following differences: (i) It takes the maximum number of signatures $J$ as an additional parameter, (ii) It generates a distinct forward-secure signature private/public key pair (step 3), (iii) It generates a private key tuple $(y^{\ell}_1,r^{\ell}_1)$ (Step 4), for each $\{\server^\ell\}_{\ell=1}^{L}$.   Unlike \lrsha, \comc~will produce  one-time public key pairs from those and certify them with a forward-secure signature.

In \flrshasig, unlike \lrsha, the signer calls a key update function \flrshaupd~(step 4), which  evolves private key pairs $\{(y^{\ell}_j,r^{\ell}_j)\}_{\ell=1}^{L}$ by hashing and then deleting the previous key given $1\leq j < J$  (steps 2-3). \flrshaupd~ensures a forward-secure private key pair is maintained per \comc~server up to state $J$. \flrshasig~then computes two aggregate private key components (step 1) instead of one, but uses this key pair as in \lrshasig~to compute the signature (steps 2-3).  The cost of \flrshasig~is mainly a few PRF calls and modular additions and therefore is highly efficient, as shown in Section \ref{sec:performance_analysis}.

\flrshaver~works like \lrshaver~but with differences in aggregate keys and \comconstr:  (i) \flrshacomconstr~provides  a pair of one-time public key set and forward-secure signatures $ (\vec{Y_j}, \vec{R_j}, \vec{C_j})$ for each $\{\server^\ell\}_{\ell=1}^{L}$, which can be retrieved and authenticated offline (before actual signature verification, only up to $J$) (steps 1-3). (ii) The verifier computes the aggregate pair  $(Y_j^{1,L} , R_j^{1,L})$ (as opposed to only aggregate commitment in \lrshaver), and the rest is as in \lrshaver. Hence, the online overhead of \flrshaver~is almost as efficient as that of \lrshaver~with only a small-constant number of (negligible) scalar addition cost differences.  

{\em $\bullet$ Enhancing computational efficiency on \comc~servers}: In Algorithm \ref{alg:flrsha}, \comc~servers run a hash chain on their private key components to generate public keys and commitments. To avoid the cost of hash recursion for long chains (e.g., $\approx 2^{20}$), one can use a pre-computed table of the private key components with interleaved indices. This offers a computation-storage trade-off that \comc~servers can decide. Note that the private keys are stored in a secure enclave. Given that modern enclaves offer large protected memory of up to $512$ GB, the overhead of pre-computed tables is likely negligible. For instance, the total memory overhead of $(J=2^{20})$ public commitments is equal to only $32$ MB. 


	\section{Security Analysis} \label{sec:SecurityAnalysis}
We prove that \lrsha~and \flrsha~are \HDEUCMA~and \FHDEUCMA~secure, respectively (in random oracle model). We omit terms negligible to $\kappa$ (unless expressed for clarity).  

%
\begin{theorem} \label{the:LRSHA}
	If a PPT adversary \A~can break the \HDEUCMA-secure \lrsha~in time $t$ and after $q_s$ signature and commitment queries to \Lrshasig~and \Lrshacomconstr~oracles, and $q_H$ queries to \ro, then one can build a polynomial-time algorithm \F~that breaks the \dlp~in time $t'$ (by Definition \ref{def:HDEUCMA}). The probability that any $\{S^l\}_{l=1}^{L}$ injects a false commitment without being detected is  \advsgnproof, under the Assumption \ref{assump:tee2},  in time $t''$. 
	\begin{eqnarray*}
		\resizebox{0.48\textwidth}{!}
		{
			$\advlrsha \le {\advdll},~t'=\bigo(t)+ L \cdot \bigo(q_s\cdot \kappa^3)$
		}
	\end{eqnarray*}
\end{theorem}

\vspace{1pt}
\noindent {\em Proof:}  
Let \A~be a \lrsha~attacker. We construct a \dl-attacker that uses \A~as a subroutine. We set $(y \Rq , Y \as \alpha^y \mod p)$ as in Definition~\ref{def:dlp} and  $\vec{\pk'} = \{ (\sk'^\ell,\pk'^\ell)\as \sgnkg(1^\kappa) \}_{\ell=1}^L$  (as in \lrshakg), where the rest of the key generation will be simulated in the {\em Setup phase} . \F~is run by Definition~\ref{def:HDEUCMA} (i.e., \HDEUCMA) as follows: 

\vspace{1mm} \noindent \underline{{\em Algorithm $\mathcal{F}(\pk)$}}:

\underline{ {\em Setup}:} 
\F~maintains \lh,\lm, and \lr~to keep track of the query results in during the experiments. 
\lh~is a public hash list in form of pairs $\{M_i:h_i\}$, where $(M_i,h_i)$ represents $i^{\text{th}}$ data item queried to \ro~and its corresponding answer, respectively. 
\lm~is a public message list that represents the messages queried by \A~to $\Lrshasig$ oracle. 
%
%
\lr~is a private list containing the randomly generated variables.

\vspace{2pt}
\F~initializes simulated public keys and \ro~as follows:

\begin{enumerate}[-]
	\item {\em Key Setup}: \F~injects challenge public key as $\pk=(Y, \vec{\pk'})$, and sets the parameters $I \as (p,q,\alpha,L)$. \F~generates $x_0 \Rq$, adds it to \lr, and sets the state  as $St \as j=1$. 
	%
	%
	\item {\em \ro~Setup}: \F~uses a function \hsim~that acts as a random oracle \ro. If $\exists M : \lh[M]=h$, then \hsim~returns $h$. Otherwise, it returns $h \Rq$ and save it as $\lh[M] \as h$. 
\end{enumerate}

\vspace{3pt}

$\bullet$ \underline{{\em Execute \resizebox{.76\hsize}{!}{$(M^{*},\sigma^{*})\as \mathcal{A}^{ \ro,~\Lrshasig, ~\Lrshacomconstr}(\pk)$} }}: 

\F~handles the queries of \A~as follows:

\begin{enumerate}[-]
	\setlength{\itemsep}{1pt}
	\setlength{\parskip}{0pt}
	\setlength{\parsep}{0pt}
	
	\item  \underline{Queries of \A}: \A~can query \ro~and $\Lrshasig$ on any message $M$ of its choice up to $q_H$ and $q_s$ times, respectively.  \A~can query \Lrshacomconstr~oracle on the state $j$ as input, and it returns the corresponding commitments $(\vec{R_j},\vec{C_j})$ as the output. \F~handles \A's queries as follows:
	
	%
	\begin{enumerate}[1)]
		\itemsep 1pt
		\item {\em \ro~queries}: \A~queries \ro~on a message $M$. \F~calls $h\as\hsim(M,\lh)$ and returns $h$ to \A.
		
		\item {\em $\Lrshasig$~queries}: Insert $M$ into \lm, and execute:
		
		
		\begin{enumerate}[i)]
			\setlength{\itemsep}{2pt}
			\setlength{\parskip}{0pt}
			\setlength{\parsep}{0pt}
			
			\item $x_j \as \hsim(x_0 \| j, \lh)$, where $(x_0 \as \lr)$. If $M_j \| x_j \notin \lh$, then \F~calls $\hsim(M_j \| x_j, \lh)$, otherwise \F~{\em aborts}.
			\item Retrieve $s_j$ from \lr~if $s_j \in \lr$. Otherwise, \F~sets $s_j \Rq$, $e_j \Rq$, and $R_j^{1,L} \as \alpha^{s_j} \cdot Y^{e_j} \mod p $ and adds each of $s_j,e_j,$ and $R_j^{1,L}$ to \lr. 
			\item  $St = (j \as j+1)$ and return $\sigma_j \as (s_j,x_j,j)$.
		\end{enumerate}
		
		\item {\em Handle $\Lrshacomconstr$}: \A~can query \Lrshacomconstr~on any index $1\leq j \leq q_s$ of her choice. \F~handles these queries as follows: If  $R_j^\ell \notin \lr$, then \F~generates $R_j^\ell \Rp$ and adds it to \lr, else fetch it from \lr~for $\ell =1, \ldots, L-1$.   \F~also checks if $R_j^{1,L} \in \lr$, then it retrieves. Otherwise, it generates $e_j \Rq$ and $s_j \Rq$, and sets $R_j^{1,L} \as \alpha^{s_j} \cdot Y^{e_j} \mod p$,  $R_j^L \as \prod_{\ell_1=1}^{L-1}{(R_{j}^{\ell_1})}^{-1} \cdot R_j^{1,L} \mod p$,  and add these values to \lr. Finally,  \F~computes $\{C_j^\ell \as \sgnsig_{\vec{\sk'}}(R_j^\ell)\}_{\ell=1}^{L}$ and returns $(\vec{R_j} \as \{R_j^\ell\}_{\ell=1}^L,\vec{C_j}=\{C_j^\ell\}_{\ell=1}^L)$ to \A. 
		
		
		
	\end{enumerate}
	\vspace{1pt}
	
	\item \underline{Forgery of~\A}: Finally, $\mathcal{A}$ outputs a forgery for \pk~as $(M^{*},\sigma^{*})$, where $\sigma^{*}=(s^{*},x^*,j)$. By Definition \ref{def:HDEUCMA}, \A~wins the \HDEUCMA-experiment for \lrsha~if $\lrshaver(\pk, M^{*},\sigma^{*})=1$ and  $M^{*} \notin \lm$. 
	%
	
	\item \underline{Forgery of \F}: If \A~fails, \F~also fails and {\em aborts}. Otherwise, given an \lrsha~forgery $(M^*,\sigma^*\as (s^*,x^*,j	))$ on \pk: {\em (i)} \F~checks if $M^* \| x^* \notin \lh$ (i.e., \A~does not query \ro), then \F~{\em aborts}. 
	{\em (ii)} \F~checks if $R_j^{1,L} \notin \lr$ (i.e., \F~did not query \Lrshacomconstr), then \F~{\em aborts}. 
	Otherwise, \F~continues as follows: Given $R_j^{1,L}$ computed by \A, the equation $R_j^{1,L}=Y^{e_j} \cdot \alpha^{s_j} \mod p$ holds, where $e_j$ and $s_j$ are derived from \lr.	$\lrshaver(\pk,M_j^*,\sigma_j^*)=1$ also holds, and so  $R_j^{1,L} \equiv Y^{e_j^*} \cdot \alpha^{s_j^*} \mod p$ holds and  $e_j^* \as \hsim(M_j^* \| x_j^*) \mod p$. 	Therefore, \F~can extract $y'=y$ by solving the below modular linear equations, where $Y=\alpha^{y'}\mod p$.  
	
	\vspace{-5mm}
	\begin{eqnarray*}
		R_j^{1,L} \equiv Y^{e_j^{*}}\cdot \alpha^{s_j^{*}}  \bmod p, & R_j^{1,L} \equiv  Y^{e_j}\cdot \alpha^{s_j} \bmod p, \\
			r_j^{1,L} \equiv y' \cdot e_j^* + s_j^* \bmod q, & r_j^{1,L} \equiv y' \cdot e_j + s_j \bmod q,
	\end{eqnarray*}
	
%
%
	
%
	
	$Y = \alpha^{y'} \mod p$ holds as \A's forgery is valid and non-trivial on $\pk$. By Definition \ref{def:dlp}, \F~wins the \dl~experiment. 
	
\end{enumerate}


\vspace{2mm}
\noindent \underline{{\em Execution Time Analysis}}: The runtime of \F~is that of \A~plus the time to respond to the queries of \ro, \Lrshasig, and \Lrshacomconstr. The dominating overhead of the simulations is modular exponentiation, whose cost is denoted as $\bigo(\kappa^3)$. Each \Lrshasig~and  \Lrshacomconstr~query invokes approximately $L$ modular exponentiation operations, making asymptotically dominant cost $L \cdot \bigo(q_s \cdot \kappa^3)$. Therefore, the approximate running time of \F~is  $t'=\bigo(t)+ L \cdot \bigo(q_s\cdot \kappa^3)$.

\vspace{2mm}
\noindent \underline{{\em Success Probability Analysis}}: 
\F~succeeds if below events occur.

\begin{enumerate}[-]
	\setlength{\itemsep}{0pt}
	\setlength{\parskip}{0pt}
	\setlength{\parsep}{0pt}
	\item \nab: \F~does not abort during the query phase.
	
	\item \forge: \A~wins the \HDEUCMA~experiment for \lrsha.
	
	\item \nabb: \F does not abort after \A's forgery.
	
	\item \suc: \F wins the \dl{\em-experiment}.
	
	\item  $Pr[\suc] = Pr[\nab]\cdot Pr[\forge|\nab]\cdot Pr[\nabb|\nab \wedge \forge]$
\end{enumerate}

\vspace{2pt}
$\bullet$ {\em The probability that event \nab~occurs}: During the query phase, \F~aborts if $M_j \| x_j \in\lh$ holds, {\em before} \F~inserts $M_j \| x_j$ into \lh. This occurs if \A~guesses $x_j$ (before it is released) and then queries $M_j \| x_j$ to \ro~{\em before} it queries $\Hdsgnsig$. The probability that this occurs is $\frac{1}{2^{\kappa}}$, which is negligible in terms of $\kappa$. Hence, $Pr[\nab]=(1-\frac{1}{2^{\kappa}})\approx 1$.

$\bullet$ {\em The probability that event \forge~occurs}: If \F~does not abort, \A~also does not abort since \A's simulated  view is {\em indistinguishable} from \A's real view. 
%
%
Therefore, $Pr[\forge|\nab] \approx \advlrsha$.

$\bullet$ {\em The probability that event \nabb~occurs}: \F~does not abort if the following conditions are satisfied:
(i) \A~wins the \HDEUCMA~experiment for \lrsha~on a message $M^{*}$ by querying it to \ro. The probability that \A~wins without querying $M^{*}$ to \ro~is as difficult as a random guess (see event \nab). 
(ii) \A~wins the \HDEUCMA~experiment for \lrsha~by querying \Lrshacomconstr. The probability that \A~wins without querying \Lrshacomconstr~is equivalent to forging \sgn, which is equal to \advsgnp. 
(iii) After \F~extracts $y'=y$ by solving modular linear equations, the probability that $Y \not\equiv \alpha^{y'} \bmod p$ is negligible in terms $\kappa$, since $\pk=(Y,\{\pk'^\ell\}_{\ell=1}^L)$ and $\lrshaver(\pk,M^{*},\sigma^{*})=1$. Hence, $Pr[\nabb|\nab \wedge \forge] \approx \advlrsha $.


\vspace{2mm}
\noindent {\em \underline{Indistinguishability Argument}}: The real-view of  \Areal~is comprised of $(\pk,I)$, the answers of $\Lrshasig$ and $\Lrshacomconstr$~(recorded in \lm~and \lr~by \F), and the answer of \ro~(recorded in \lh~by \F).  All these values are generated by \lrsha~algorithms, where $\sk=( y,\vec{r} )$ serves as initial randomness. The joint probability distribution of \Areal~is random uniform as that of \sk. 

The simulated view of \A~is as \Asim, and it is equivalent to \Areal~except that in the simulation, $(s_j,e_j,x_0)$ (and so as $x_j)$ are randomly drawn from $\mathbb{Z}_q^*$. 
This dictates the selection $\{R_j^{1,L}\}$ as random via the $\Lrshasig$~and $\Lrshacomconstr$~oracles, respectively. That is,
for each state $St=j$, the partial public commitments $\{R_j^\ell\}_{\ell=1}^{L-1}$ are randomly selected from $\mathbb{Z}_q^*$ while the last partial public commitment is equal to $R_j^L \as \prod_{\ell=1}^{L-1}{(R_{j}^{\ell})}^{-1} \cdot \alpha^{s_j} \cdot Y^{e_j} \mod p$.
The aggregate commitment $R_j^{1,L} \as \prod_{\ell=1}^{L} R_j^\ell \mod p = (\prod_{\ell=1}^{L-1} R_j^\ell) \cdot (\prod_{\ell=1}^{L-1}{(R_{j}^{\ell})}^{-1}) \cdot \alpha^{s_j} \cdot Y^{e_j} = \alpha^{s_j} \cdot Y^{e_j} \mod p $. 
Thus, the correctness of aggregate commitments holds as in the real view.
The joint probability distribution of these values is randomly and uniformly distributed and is identical to original signatures and hash outputs in \Areal, since the cryptographic hash function $H$ is modeled as \ro~via \hsim. 
\hfill $\blacksquare$


\begin{figure*}[!h]
	\centering
	\begin{tikzpicture}[
		node distance=2.9cm,
		every node/.style={font=\sffamily},
		process/.style={rectangle, rounded corners, draw=black!70, thick, fill=gray!8, text width=3.6cm, align=center, minimum height=5.5cm, inner sep=5pt},
		innerbox/.style={rectangle, rounded corners, draw=black!60, fill=white, text width=3.0cm, align=center, minimum height=0.6cm, font=\footnotesize},
		arrow/.style={->, >=Stealth, thick}
		]
		
		\node[process, fill=green!10] (signer) {};
		\node[innerbox, below=0.1cm of signer.north] (signer1) {\mr{Finite-field Operations}\\\mr{($\text{Add}_q$, $\text{Sub}_q$, $\text{Mul}_q$)}};
		\node[innerbox, below=0.1cm of signer1] (signer2) {\mr{Cryptographic Hash Function}\\\mr{(Ascon-Hash-256)}};
		\node[above=0.0cm of signer.south, font=\bfseries] (signerlabel) {\mr{Signer (IoT Device)}};
		
		\node[process, fill=blue!10, right=of signer] (verifier) {};
		\node[innerbox, below=0.1cm of verifier.north] (ver1) {\mr{Finite-field Operations}\\\mr{(e.g., $\text{Add}_q$, $\text{Sub}_q$, $\text{Mul}_q$)}};
		\node[innerbox, below=0.4cm of ver1] (ver2) {\mr{EC Operations}\\\mr{(e.g., EC scalar multiplication)}};
		\node[innerbox, below=0.1cm of ver2] (ver3) {\mr{Cryptographic Hash Function}\\\mr{(Ascon-Hash-256)}};
		\node[above=0.0cm of verifier.south, font=\bfseries] (verlabel) {\mr{Verifier}};
		
		\node[process, fill=orange!10, right=of verifier] (comc) {};
		\node[innerbox, below=0.1cm of comc.north] (comc1) {\mr{Finite-field Operations}\\\mr{(e.g., $\text{Add}_q$, $\text{Sub}_q$, $\text{Mul}_q$)}};
		\node[innerbox, below=0.3cm of comc1] (comc2) {\mr{EC Operations}\\\mr{(EC scalar multiplication)}};
		\node[innerbox, below=0.1cm of comc2] (comc3) {\mr{Cryptographic Hash Function}\\\mr{(Ascon-Hash-256)}};
		\node[innerbox, below=0.1cm of comc3] (comc4) {\mr{Generic Signature}\\\mr{Scheme (SGN)}};
		\node[above=0.0cm of comc.south, font=\bfseries] (comclabel) {\mr{ComC Server $\boldsymbol{\server_\ell}$}};
		
		\draw[arrow] (signer) -- node[above, yshift=0.1cm]{\mr{$(M_j, \sigma_j)$}} (verifier);
		\draw[arrow] (comc) -- node[above, yshift=0.1cm]{\mr{$(R_\ell, C_\ell)$}} (verifier);
		\draw[arrow, dashed] (comc.south) .. controls +(0,-1.0cm) and +(0,-1.0cm) .. node[above]{\scriptsize \mr{Certified Commitments}} (verifier.south);
		\draw[arrow] (comc1) -- node[above, yshift=0.1cm]{} (comc2);
		\draw[arrow] (ver1) -- node[above, yshift=0.1cm]{} (ver2);
		
	\end{tikzpicture}
	\caption{\mr{High-level overview of the (\texttt{F})\lrsha~building blocks and their interplay. Each entity’s internal components show the main cryptographic and arithmetic methods used to implement each of the signature algorithms.}}
	\label{fig:lrsha_components_bottom}
\end{figure*}

\begin{theorem} \label{the:FLRSHA}
	If a PPT adversary \A~can break the \FHDEUCMA-secure \flrsha~in time $t$ and after $q_s$ signature and commitment queries to both  \Flrshasig~and \Flrshacomconstr~oracles, $q_H$ queries to \ro~and one query to \Breakin~oracle, then one can build a polynomial-time algorithm \F~that breaks \dlp~in time $t'$ (by Definition \ref{def:FHDEUCMA}).  The probability that $\{S^\ell\}_{\ell=1}^{L}$ injects a false commitment without being detected is  \advfsgnproof, under the Assumption \ref{assump:tee2},  in time $t''$.
	\begin{eqnarray*}
		\resizebox{0.48\textwidth}{!}
		{
			$\advflrsha \le {\advdll},  t'=\bigo(t)+L \cdot \bigo(J \cdot \kappa^3)$
		}
	\end{eqnarray*}
\end{theorem}

\vspace{1pt}
\noindent {\em Proof:}  Let \A~be a \flrsha~attacker and $(y \Rq, Y \as \alpha^y \mod p)$ be a \dlp~challenge as in Definition \ref{def:dlp}. We set the certification keys ($\{\sk'^\ell,\pk'^\ell\}_{\ell=1}^L$ via \fsgnkg. We then run the simulator \F~by Definition~\ref{def.hfsgn}: 


\vspace{3pt}
$\bullet$ {\em Setup:} \F~manages the lists \lm, \lh, and \lr, and handles \ro~queries via \hsim~as in Theorem \ref{the:LRSHA}. Per Definition \ref{def:FEUCMA}, \F~selects the maximum number of signatures as $J$, and then selects an index $w \Ra [1,J]$ hoping that \A~will output his forgery on. \F~continue as follows:



$-$ \underline{\em \sk~Simulation}:  Set $\sk_1=\langle \vec{y}_1 , \vec{r}_1 \rangle$ as in \flrshakg. 

	\begin{enumerate}[1:]
		\item $a^\ell \as \langle y_1^\ell,r_1^\ell, \sk'^\ell \rangle, ~\forall \ell =1,\ldots,L $, and $\vec{a}=\{a^\ell\}_{\ell=1}^L$
		\item $\sk_{j+1} \as \flrshaupd(\sk_j), \forall j \in [1,w-2]$, where $H$ in \flrshaupd~is simulated via \hsim. 
		\item $\sk_{w+1}=\langle \vec{y}_{w+1} , \vec{r}_{w+1} \rangle$ as in \flrshakg. 
		\item $\sk_{j+1} \as \flrshaupd(\sk_j), \forall j \in [w+1,J-1]$. 
		\item Add $\sk_j$ to \lr, $\forall j \in [1,w-1] \cup [w+1,J] $ and $\vec{a}$ to \lr.
	\end{enumerate}

$-$ \underline{\em \pk~and \comc~Simulation}: 
	\begin{enumerate}[1:]
		\item Retreive $\{ \sk_j \}_{ j \in [1,w-1] \cup [w+1,J] }$ and $\vec{a}$ from \lr. 
		\item \textbf{for}  $ j \in [1,w-1] \cup [w+1,J] $ \textbf{do}
		\item ~~~~~$\vec{R}_j \as \{ R_j^\ell  \as  \alpha^{ r_j^\ell } \mod q \}_{\ell=1}^L $ and $\vec{Y}_j \as \{ Y_j^\ell \as  \alpha^{ y_j^\ell } \mod q \}_{\ell=1}^L $. \F~adds $\vec{Y}_j$ and $\vec{R}_j$ to \lr.
		\item $s_w \Rq$ and $e_w \Rq$. \F~adds $s_w$ and $e_w$ to \lr.
		\item $Y_w^\ell \Rp$ and $R_w^\ell \Rq ~, \forall \ell =1,\ldots L-1 $.
		\item $Y_w^L \as  Y \cdot \prod_{\ell=1}^{L-1} {(Y_w^\ell)^{-1}} \mod p $ 
		\item $R_w^L \as  Y^{e_w} \cdot \alpha^{s_w} \cdot \prod_{\ell=1}^{L-1} ({R_w^\ell})^{-1} \mod p $ 
		\item Add $\vec{Y}_w \as \{Y_w^\ell\}_{\ell=1}^L$ and $\vec{R}_w \as \{R_w^\ell\}_{\ell=1}^L$ to \lr. 
	\end{enumerate}


\vspace{3pt}
$\bullet$ {\em Query Phase:}  \F~handles \ro~and \Breakin~queries as in  Theorem~\ref{the:LRSHA} and Definition \ref{def:FHDEUCMA}, respectively. The rest of \A's queries are as follows:


	$-$ \underline{$\Flrshasig$}:  If $j \neq w$ , then it retrieves $\sk_j$ from \lr~and computes $\sigma_j$ as in $\flrshasig$. Otherwise,  it retrieves $s_w$ from \lr~and returns $\sigma_j=(s_w,x_w \Ra \{0,1\}^\kappa ,w)$. 
	
	
	$-$ \underline{$\Flrshacomconstr$}:  \F~retrieves $(\vec{Y}_j,\vec{R}_j)$ from  \lr~and computes  $ \vec{C}_j = \{ ~C_j^\ell \as \fsgnsig_{\vec{\sk'}}(Y_j^\ell , R_j^\ell)~ \}_{\ell=1}^L$. Finally, \F~returns $(\vec{Y_j},\vec{R_j},\vec{C_j})$ to \A. 




\vspace{3pt}
$\bullet$ {\em Forgery and Extraction:} \A~outputs a forgery on \pk~as $(M^*,\sigma^*)$, where $\sigma^*=(s^*,x^*,j)$. By Definition ~\ref{def:FHDEUCMA}, \A~wins if $\flrshaver(\pk,M^*,\sigma^*)=1$, and $M^* \notin \lm$. 
\F~wins if (i) \A~wins, (ii) \A~produces a forgery on $j=w$ by querying  $\Flrshacomconstr$ (i.e.,  $(\vec{Y}_w,\vec{R}_w) \in \lr$) and \ro~(i.e., $(M^* \| x^*) \in \lh$).  If these conditions hold, then \F~extracts $y'$ as in Theorem \ref{the:LRSHA} extraction phase. 


\vspace{3pt}
$\bullet$ {\em Success Probability and Execution Time Analysis:} The analysis is similar to Theorem \ref{the:LRSHA} except the forgery index must be on $j=w$.  That is, the probability that \A~wins \FHDEUCMA~experiment against \flrsha~is equal to \advflrsha. 
\F~wins the \dlp~experiment if \A~outputs his forgery on $j=w$. Since $w$ is randomly drawn from $[1,J]$, the probability that \A~returns his forgery on $j=w$ is $1/J$. The probability that \A~wins the experiment without querying \ro~and \Lrshacomconstr~are $1/2^{\kappa}$ and \advfsgnproof, respectively. The execution time is asymptotically similar to that of Theorem \ref{the:LRSHA}, where $q_s=J$: 

%
\vspace{-8pt}
\begin{eqnarray*}
	\resizebox{0.48\textwidth}{!}
	{
			$\advflrsha \le {\advdll},  t'=\bigo(t)+2L\cdot J \cdot \bigo(\kappa^3)$
	}
\end{eqnarray*}

\vspace{-3pt}
$\bullet$ {\em Indistinguishability Argument:}  \A's real $\mathcal{A}_R$ and simulated $\mathcal{A}_S$ views are indistinguishable. The argument is as in Theorem \ref{the:LRSHA}, with the following differences: (i) In $\mathcal{A}_S$, the simulator uses private keys that are randomly generated except during $j=w$ where he injects the challenge $Y$. These random variables are identical to $\mathcal{A}_R$ since $H$ is a random oracle.  (ii) \fsgn~is used to sign commitments in the transcripts instead of \sgn. \hfill $\blacksquare$


\begin{table*}[ht!]
	\caption{Performance comparison of \lrsha~and \flrsha~schemes and their counterparts on commodity hardware}\label{tab:eses_com_hrdwr1}
	\centering
	
		\resizebox{0.98\textwidth}{!}{
			\begin{tabular}{| l || @{}c@{} | @{}c@{} | @{}c@{} | @{}c@{} | @{}c@{} || @{}c@{} | @{}c@{} || @{}c@{} | @{}c@{} | @{}c@{} | @{}c@{} | @{}c@{} | @{}c@{} | @{}c@{} }
				\hline
				
				\multirow{2}{*}{\textbf{Scheme}} 
				& \multicolumn{5}{c||}{ \textbf{Signer} } 
				& \multicolumn{2}{c||}{ \textbf{Verifier} } 
				& \multicolumn{6}{c|}{\textbf{\comc~Servers}} 
				\\ \cline{2-14}
				
				& \multirow{2}{*}{ \specialcell[]{ \textbf{Signing} \\ \textbf{Time ($\mu$s)} }  }
				& \multirow{2}{*}{ \specialcell[]{ \textbf{Private} \\ \textbf{Key (KB)} }   }
				& \multirow{2}{*}{ \specialcell[]{ \textbf{Signature} \\ \textbf{Size (KB)} } }
				& \multirow{2}{*}{ \specialcell[]{ \textbf{Forward} \\ \textbf{Security} } }
				& \multirow{2}{*}{ \specialcell[]{ \textbf{Online} \\ \textbf{Sampling} } }
				& \multirow{2}{*}{ \specialcell[]{ \textbf{Public} \\ \textbf{Key (KB)} } }
				& \multirow{2}{*}{ \specialcell[]{ \textbf{Ver} \\ \textbf{Time ($\mu$s)}} }
				& \multirow{2}{*}{ \specialcell[]{ \textbf{Storage Per} \\ \textbf{Server (KB)} } }
				& \multicolumn{2}{c|}{ \specialcell[]{ \textbf{Comp. Per} \\ \textbf{Server} } }
				& \multirow{2}{*}{ \specialcell[]{ \textbf{Collusion} \\ \textbf{Resiliency} } }
				& \multirow{2}{*}{ \specialcell[]{ \textbf{False} \\ \textbf{Input} \\ \textbf{Detection} } }
				& \multirow{2}{*}{ \specialcell[]{ \textbf{Offline} \\ \textbf{Gen.} } }
				\\ \cline{10-11}
				& 
				& 
				&
				& 
				& 
				& 
				& 
				& 
				& \specialcell[]{ \textbf{Key} \\ \textbf{Gen.} }
				& \specialcell[]{ \textbf{Cert.} \\ \textbf{Gen.} }
				& 
				&
				&
				
				\\ \hline
				
				ECDSA~\cite{ECDSA} & $17.07$ & $0.03$  & $0.06$ & $\times$ & $\checkmark$ & $0.03$ & $46.62$ & \multicolumn{6}{c|}{}  \\  \cline{1-8}
				Ed25519~\cite{Ed25519} & $16.34$ & $0.03$  & $0.06$ & $\times$ & $\checkmark$ & $0.03$ & $39.68$ &  \multicolumn{6}{c|}{ } \\ \cline{1-8}
				Ed25519-BPV~\cite{Ed25519} & $19.96$ & $1.03$  & $0.06$ & $\times$ & $\checkmark$ & $0.03$ & $39.68$ & \multicolumn{6}{c|}{ }  \\ \cline{1-8}
				Ed25519-MMM~\cite{Ed25519} & $82.32$ & $53.09$ & $1.2$ & $\checkmark$ & $\checkmark$ & $0.03$ & $267.04$ &  \multicolumn{6}{c|}{ } \\ \cline{1-8}
				BLS~\cite{BLS:2004:Boneh:JournalofCrypto} & $278.6$ & $0.06$  & $0.05$ & $\times$ & $\checkmark$ & $0.09$ & $910.6$ &  \multicolumn{6}{c|}{ }  \\ \cline{1-8}
				RSA-3072~\cite{seo2020efficient} & $1235.74$ & $0.5$  & $0.25$ & $\times$ & $\checkmark$ & $0.5$ & $45.78$ &  \multicolumn{6}{c|}{ N/A }  \\ \cline{1-8}
				
				SCRA-BLS~\cite{SCRA:Yavuz} & $15.31$ & $16.06$  & $0.05$ & $\times$ & $\times$ & $0.09$ & $43.52$ &  \multicolumn{6}{c|}{ }  \\ \cline{1-8}
				SCRA-RSA~\cite{SCRA:Yavuz} & $22.99$ & $2$ MB  & $0.27$ & $\times$ & $\times$ & $0.53$ & $51.2$ &  \multicolumn{6}{c|}{ }  \\ \cline{1-8}
				
				BLISS-\romannum{1}~\cite{BLISS} & $241.3$  & $2.00$ & $5.6$ & $\times$ & $\checkmark$ & $7.00$ & $24.61$ &  \multicolumn{6}{c|}{ }  \\ \cline{1-8}
				SPHINCS+~\cite{bernstein2019sphincs+} & $5,445.2$ & $0.1$  & $35.66$ & $\times$ & $\times$ & $0.05$ & $536.14$ &  \multicolumn{6}{c|}{ }  \\ \cline{1-8}
				\xmssmt~\cite{cooper2020recommendation} & $10,682.35$  & $5.86$ & $4.85$ & $\checkmark$ & $\times$ & $0.06$ & $2,098.84$ &  \multicolumn{6}{c|}{ }  \\ \hline
				
				$\text{HASES}$~\cite{Yavuz:HASES:ICC2023} & $5.89$ & $0.03$  & $0.5$ & $\checkmark$ & $\times$ & $32$ & $10.41$ & $32$ & $624.64~\mu s$  & N/A  & Central & $\times$ &  $\checkmark$  \\ \hline 
				
				$\text{ESEM}_2$~\cite{Yavuz:CNS:2019} & $10.34$ & $12.03$  & $0.05$ & $\times$ & $\times$ & $0.03$ & $259.79$ & $4.03$ & $82.91~\mu s$ & N/A & Semi-Honest & $\times$ & $\times$ \\ \hline \hline
				
				LRSHA & $\boldsymbol{3.23}$ & $\boldsymbol{0.06}$  & $\boldsymbol{0.05}$ & $\times$ & $\times$ & $\boldsymbol{0.03}$ & $\boldsymbol{45.96}$ & $\textbf{0.03}$ & $\boldsymbol{2.56~ms}$ & $\boldsymbol{22.67~\mu s}$ &  \textbf{Protected} & $\checkmark$ & $\checkmark$  \\ \hline
				
				FLRSHA & $\boldsymbol{5.35}$ & $\boldsymbol{0.22}$ & $\boldsymbol{0.05}$ & $\checkmark$ & $\times$ & $\boldsymbol{0.03}$ & $\boldsymbol{45.96}$ & $\textbf{0.06}$ & $\boldsymbol{5.53~ms}$ & $\boldsymbol{82.32~\mu s}$ &  \textbf{Protected} & $\checkmark$ & $\checkmark$  \\ \hline
				
			\end{tabular}
		}
		
		\begin{tablenotes}[flushleft] 
			\scriptsize{
				\item The input message size is 32 bytes. The maximum number of signing for FS schemes is set to $J=2^{20}$. The number of  \comc~servers is $L=3$. 
				For SPHINICS+ parameters, $n=16,~h=66,~d=22,~b=6,~k=33,~w=16$ and $\kappa=128$.
				We benchmark the XMSSMT\_SHA2\_20\_256 variant, allowing for $2^{20}$ signings.
				HASES parameters are $(l=256,t=1024,k=16)$. BPV parameters in $\text{ESEM}_2$ are $(n=128,v=40)$.
				The parameter $n$ in RSA is $3072$-bit. 
				For SCRA-BLS and SCRA-RSA, we set the optimal setting $(L=32, b=8)$.
				The online verification time of \lrsha~and \flrsha~is similar to that of Ed25519. The (offline) aggregation of commitments for \lrsha~and \flrsha~is $9.1~\mu s$ and $16.99~\mu s$, respectively.
				However, in $\text{ESEM}_2$, the verification includes the commitment aggregation (i.e., $R$). Indeed, the \comc~servers in ESEM require verifier input before generating the commitments. One can delegate the aggregation of partial commitments to a \comc~server, but it will incur more network delay. 
			}
		\end{tablenotes}
	\end{table*}

	\begin{table*}[ht!]
		\caption{Performance comparison of \lrsha~and \flrsha~schemes and their counterparts on 8-bit AVR ATmega2560 MCU}\label{tab:eses_8bit}
		\centering
		
		\resizebox{0.8\textwidth}{!}{
			\begin{tabular}{| l || c | c | c | c | c | c | @{}c@{} | @{}c@{} | @{}c@{} | @{}c@{} | }
				\hline
				\textbf{Scheme} 
				& \specialcell[]{ \textbf{Signing} \\ \textbf{(Cycles)} }  
				& \specialcell[]{ \textbf{Signing/Sensor} \\ \textbf{Energy Ratio (\%)} }  
				& \specialcell[]{ \textbf{Private} \\ \textbf{Key (KB)} }  
				& \specialcell[]{ \textbf{Signature}  \\ \textbf{Size (KB)} }  
				& \specialcell[]{ \textbf{Forward} \\ \textbf{Security} } 
				& \specialcell[]{\textbf{Precomputation} \\ \textbf{Feasibility}}
				& \specialcell[]{\textbf{Simple Code} \\ \textbf{Base}}
				\\ \hline \hline
				
				ECDSA~\cite{ECDSA} & $79,185,664$ & $93.75$ & $0.03$  & $0.05$  & $\times$  & $\times$ & $\times$ \\ \hline
				Ed25519~\cite{Ed25519} & $22,688,583$ & $26.86$ & $0.03$  & $0.06$  & $\times$  & $\times$ & $\times$  \\ \hline			
				BLISS-\romannum{1}~\cite{BLISS} & $10,537,981$  & $12.48$ & $2.00$ & $5.6$ & $\times$ & $\times$  & $\times$  \\ \hline
				HASES \cite{Yavuz:HASES:ICC2023} & $1,974,528$ & $2.34$ & $0.05$  & $0.5$  & $\checkmark$  & $\times$  & $\checkmark$  \\ \hline
				$\text{ESEM}_2$~\cite{Yavuz:CNS:2019} & $1,555,380$ & $1.84$ & $12.03$  & $0.05$  & $\times$  & $\checkmark$  & $\checkmark$  \\ \hline \hline
				
				LRSHA & $\boldsymbol{498,317}$ & $\boldsymbol{0.59}$ &  $\boldsymbol{0.06}$  & $\boldsymbol{0.03}$  & $\times$  & $\checkmark$  & $\checkmark$  \\ \hline
				FLRSHA & $\boldsymbol{1,602,749}$ & $\boldsymbol{1.9}$ & $\boldsymbol{0.22}$ & $\boldsymbol{0.06}$  & $\checkmark$  & $\checkmark$  & $\checkmark$   \\ \hline
			\end{tabular}
		}
		
		\begin{tablenotes}[flushleft] \scriptsize{
				\item The input message size is 32 Bytes. $\text{ESEM}_2$ incur a storage penalty of 12 KB at the signer side. The HORS parameters of HASES are $(l=256,t=1024,k=16)$. 
			}
		\end{tablenotes}
	\end{table*}

	\begin{table*}[ht!]
		\caption{Signing efficiency of \lrsha~and \flrsha~schemes via offline-online technique}\label{tab:signing_bench}
		\centering
		
		\resizebox{0.98\textwidth}{!}{			
			\begin{tabular}{| l || c | c | c | c || c | c | c | c || c | c | c | c | c | c | c | }
				\hline
				
				& \multicolumn{4}{c||}{ \textbf{Commodity Hardware} (in $\mu s$) }
				& \multicolumn{4}{c||}{ \textbf{8-bit AVR ATmega2560 MCU} (in Cycles) }
				& \multirow{3}{*} { \specialcell[]{\textbf{Additional} \\ \textbf{Storage} \\ \textbf{Cost} \\ \textbf{(KB) $^\star$}} } 
				& \multirow{3}{*} { \specialcell[]{\textbf{Forward} \\ \textbf{Security} } } 
				\\ \cline{2-9}
				
				\multirow{2}{*}{\textbf{Scheme}} 
				& \multicolumn{3}{c|}{ \textbf{Offline Computation} } 
				& \multicolumn{1}{c||}{ \textbf{Online Computation} } 
				& \multicolumn{3}{c|}{ \textbf{Offline Computation} } 
				& \multicolumn{1}{c||}{ \textbf{Online Computation} } 
				&
				&
				\\ \cline{2-9} 
				& \specialcell[]{\textbf{Priv. Key} \\ \textbf{Comp.}}  
				&  \specialcell[]{\textbf{Priv. Key} \\ \textbf{Upd.}}  
				& \textbf{Total} 
				& \textbf{Signing} 
				& \specialcell[]{\textbf{Priv. Key} \\ \textbf{Comp.}}  
				&  \specialcell[]{\textbf{Priv. Key} \\ \textbf{Upd.}}  
				& \textbf{Total} 
				& \textbf{Signing} 
				&
				&
				\\ \hline
				$\text{ESEM}_2$~\cite{Yavuz:CNS:2019} 
				& $6.25$  
				& $0$  
				& $6.25$ 
				& $4.09$ 
				& $1,006,144$ 
				& $0$  
				& $1,006,144$  
				& $549,236$ 
				& $96$
				& $\times$ \\ \hline
				Ed25519-BPV \cite{Yavuz:Dronecrypt:2018} 
				& $17.82$ 
				& $0$  
				& $17.82$  
				& $2.14$ 
				& $298,880$  
				& $0$  
				& $298,880$ 
				& $549,236$ 
				& $64$  
				&  $\times$
				\\ \hline
				LRSHA 
				& $1.83$ 
				& $0$  
				& $1.83$ 
				& $\boldsymbol{1.4}$ 
				& $376,240$ 
				& $0$  
				& $376,240$ 
				& $\boldsymbol{121,949}$ 
				& $64$ 
				&  $\times$
				\\ \hline
				
				FLRSHA 
				& $2.04$ 
				& $1.95$  
				& $3.99$ 
				& $\boldsymbol{1.36}$ 
				& $712,800$ 
				& $767,872$ 
				& $1,480,672$ 
				& $\boldsymbol{121,949}$ 
				& $128$ 
				& $\checkmark$ \\ \hline 
			\end{tabular}
		}
		
		\begin{tablenotes}[flushleft] 
			\scriptsize{
				\item The running time is in $\mu s$ for commodity hardware and in cycles for 8-bit AVR ATmega2560. The input message size is 32 Bytes. The number of \comc~servers (i.e., $L$) is $3$.
				The offline computation requires a storage penalty to save the computed keys in memory. 
				SCRA-BLS and SCRA-RSA are not present due to the large private key size and expensive signing, respectively, compared to that of Ed25519-BPV, as in Table \ref{tab:eses_com_hrdwr1}. SCRA-BLS and SCRA-RSA perform $L$ EC point additions over a gap group and $L$ modular multiplications over a large modulus (e.g., $n$ is $3072$-bit), respectively. Consequently, we considered Ed25519-BPV as most efficient OO digital signature. 
				\item $\star$ \lrsha~and \flrsha~require replenishment of additional stored data each $2^{11}$ signings. For a total of $2^{20}$ signings, replenishment of additional data is $512$ times. 
			}
		\end{tablenotes}
	\end{table*}

	\section{Performance Analysis} \label{sec:performance_analysis}
We present a comprehensive performance evaluation of our schemes with a detailed comparison with their counterparts.

	\subsection{Evaluation Metrics and Experimental Setup}
	\underline{\em Evaluation Metrics:}	We compare our proposed schemes and their counterparts based on: (1) signing computational overhead,  (2) signature size, (3) private key size (including pre-computed tables), (4) verification overhead, (5) public key size, (6) performance in pre-computed, offline/online settings, (7) forward-security, (8) collusion resiliency / false input detection, (9) implementation features (e.g., online sampling, code base simplicity), (10) impact on battery life.

	\noindent \underline{\em Selection Rationale of Counterparts:} We follow our related work analysis in Section \ref{subsec:RelatedWork} as the guideline. Given that it is not possible to compare our schemes with every single digital signature, we focus on the most relevant categories to our work, especially the ones having an open-source implementation on low-end devices: (i) ECDSA \cite{ECDSA}, BLS \cite{BLS:2004:Boneh:JournalofCrypto}, RSA \cite{seo2020efficient} to cover the most prominent signatures, serving as a building block for others. (ii) BLISS  as the lattice-based (due to its ability to run on an 8-bit MCU), SPHINCS+\cite{bernstein2019sphincs+} (hash-based on commodity hardware), and  \xmssmt~\cite{cooper2020recommendation} as the forward-secure standard. (iii) Alternative lightweight signatures with TEE and/or cloud assistance (e.g., \cite{Yavuz:HASES:ICC2023,Yavuz:CNS:2019}). (iv) We compare our forward-secure \flrsha~with MMM transformed versions of the most efficient signature schemes since it is proven to be an asymptotically optimal generic forward-secure transformation. We also compared \flrsha~with the most recent hardware-assisted counterparts~\cite{Yavuz:HASES:ICC2023}. (v) We also provided a comprehensive comparison when various signer optimizations are considered, especially with pre-computation methods for low-end devices (e.g., SCRA~\cite{SCRA:Yavuz}, BPV-variants, offline-online, etc.). 	We also included the pre-computed version $\text{ESEM}_2$~\cite{Yavuz:CNS:2019} as an ECDLP-based cloud-assisted digital signature with distributed verification. 

	\noindent \underline{\em Parameter Selection:}
	We selected the security parameter as $\kappa=128$ and ASCON-Hasha \cite{dobraunig2021ascon} as our cryptographic hash function $H$. We used the Curve25519 \cite{Ed25519} (as NIST's FIPS 186-5 standard, 256-bit public keys) for our signature schemes. For BLS \cite{BLS:2004:Boneh:JournalofCrypto}, we selected the curve BLS12-381, having an embedding degree equal to 12 and a 381 bit length. 
	We selected \xmssmt~which allows $J=2^{20}$ messages to be signed. We also set an equal signing capability $J=2^{20}$ in our schemes, \lrsha~and \flrsha. 
	We selected Ed25519 as our standard signature scheme $\sgn$ that serves to certify the commitments of \lrsha, while we opted for Ed25519 with optimal generic MMM \cite{ForwardSecure_MMM_02} as a forward-secure \fsgn~for the \comc~certification of \flrsha. We discuss the parameters and specifics of other counterparts in Table 2. 
	
	\mr{
	\noindent \underline{\em Hardware/Software Configurations:} 
	We benchmark our proposed schemes on a high-performance workstation and highly constrained 8-bit MCU as follows:
	(i) We used a desktop with an Intel i9-9900K@3.6 GHz processor and 64 GB of RAM. We also used  ASCON\footnote{\url{https://github.com/ascon/ascon-c}}, OpenSSL\footnote{\url{https://github.com/openssl/openssl}} and Intel SGX SSL\footnote{\url{https://github.com/intel/intel-sgx-ssl}} open-source libraries.
	(ii) We fully implemented \lrsha~schemes on an 8-bit AVR ATmega2560 Micro-Controller Unit (MCU) at the signer side. This MCU is an 8-bit ATmega2560, having 256KB flash memory, 8KB SRAM, and 4KB EEPROM operating at a clock frequency of 16MHz. We used $\mu$NaCl\footnote{\url{https://munacl.cryptojedi.org/atmega.shtml}} open-source software library to implement the finite-field arithmetic operations and ASCON open-source software for cryptographic hashing. }


		\mr{
		\noindent \underline{\em Implementation Details:}
		Our implementation of the (\texttt{F})\lrsha~schemes follows the architecture illustrated in Figure \ref{fig:lrsha_components_bottom}. Each entity (i.e., the Signer, Verifier, and the set of ComC servers $\{\server_\ell\}$) is realized using a distinct combination of optimized arithmetic and cryptographic primitives.
		Our {\em Signer} is designed for extremely resource-constrained IoT devices and therefore relies solely on \emph{finite-field arithmetic} and the lightweight cryptographic hash function \emph{Ascon-Hash-256}. The modular operations $(\text{Add}_q, \text{Sub}_q, \text{Mul}_q)$ are performed over a prime field $\mathbb{F}_q$ (where $q=2^{255}-19$). The signer avoids any ExpOps including elliptic-curve (EC) or modular exponentiation operations. 
		The {\em Verifier} checks a signature’s validity where it performs EC-based validation (e.g., $\alpha^s Y^e \bmod p$), the verifier executes only double scalar multiplications and hash evaluations, which are widely available in standard cryptographic libraries. 
		Each commitment constructor server $\{\server_\ell\}$ maintains its own share $r_\ell$ and produces certified commitments $(R_\ell, C_\ell)$ via one EC scalar multiplication and a digital signature \sgn~that are later verified and aggregated by the verifier. 
		Overall, the implementation of (\texttt{F})\lrsha~is backward-compatible with standard-compliant cryptographic libraries where finite-field arithmetic operations and the presence of a cryptographic hash function are available. 
		We note that we open-source our implementation for reproducibility at \url{https://github.com/saifnouma/lrsha}.
	}

	\subsection{Performance on Commodity Hardware}
	We used a desktop with an Intel i9-9900K@3.6 GHz processor and 64 GB of RAM. We also used  ASCON\footnote{\url{https://github.com/ascon/ascon-c}}, OpenSSL\footnote{\url{https://github.com/openssl/openssl}} and Intel SGX SSL\footnote{\url{https://github.com/intel/intel-sgx-ssl}} open-source libraries. 	Table \ref{tab:eses_com_hrdwr1} illustrates the overall performance of \lrsha~and their counterparts at the signer and verifier. Our main takeaways are as follows:

	$\bullet$ {\em Signing Time:} \lrsha~is $5\times$ and $3.2\times$ faster than $\text{ESEM}_2$ and standard Ed25519, respectively, but with a much smaller memory footprint than $\text{ESEM}_2$. \flrsha~offer forward security with only $1.65\times$ decrease in speedup compared to \lrsha. Notably, \flrsha~is significantly faster than its forward-secure counterpart, \xmssmt~by {\em several orders of magnitude}.  HASES~is also post-quantum-secure but suffers from a central root of trust that depends on a single hardware-supported \comc~server in order to distribute large-sized public keys to verifiers. In contrast, \flrsha~distinguishes itself by employing a network of distributed \comc~servers, thereby mitigating the risk associated with a single point of failure. Moreover, \flrsha~has {\em a magnitude times smaller signature} than that of HASES. 
	
	
	%
	%
	%

	$\bullet$ {\em Signer Storage:} \lrsha~consumes $200\times$ less memory on resource-constrained signers compared to $\text{ESEM}_2$. This is attributed to the fact that ESEM stores a set of precomputed commitments to enhance signing efficiency. 
	\flrsha~also outperforms the {\em optimal} generic forward-secure Ed25519-MMM and \xmssmt~by having $241\times$ and $27\times$ lesser memory usage, respectively. \flrsha~consumes $7\times$ more memory usage than its signer-efficient counterpart HASES, but without having a central root of trust on the \comc~servers. Additionally, \lrsha~schemes avoid costly EC-based operations by only executing simple arithmetic and symmetric operations. 
	
	$\bullet$ {\em Signature Size:} \lrsha~has the smallest signature size among all counters, with a significant signing computational efficiency at its side. Only BLS have slightly larger (i.e., $1.5\times$) signature size, while its signing operates at an order of magnitude slower pace compared to our scheme. 
	Similarly, \flrsha~surpasses its most signer-efficient forward-secure counterpart HASES with a $8.33\times$ smaller signature. Consequently, \lrsha~schemes prove to be the most resource-efficient in terms of processing, memory, and bandwidth.
	
	$\bullet$ {\em Verification Time:} We consider: (i) computation of commitments at the \comc~servers, (ii) network delay to transmit commitments and their signatures to the verifier, (iii) signature verification time at the verifier.

	
Unlike some other alternatives with distributed server support (e.g.,~\cite{Yavuz:CNS:2019}), our schemes permit an offline pre-computation of the commitments at \comc~servers. This significantly reduces the verification delay to a mere $46\mu s$ for both \lrsha~and \flrsha. Moreover, verifiers may request public commitments in batches by sending a set of counters. \flrsha's verification is slower than our fastest forward-secure hardware-assisted counterpart, HASES. However, in return, \flrsha~offers a magnitude of smaller signature sizes, faster signing, and resiliency against single-point failures. 
	
	
$\bullet$ {\em Enhanced Security Properties:} We demonstrated that \flrsha~offers a superior performance trade-off. Our most efficient counterparts assume semi-honest and non-colluding server(s), are not forward-secure, and do not authenticate the commitment. In contrast, (i) \lrsha~is forward-secure, (ii) leverages a set of SGX-supported \comc~servers to ensure resiliency against collusions and single-point failures, (iii) authenticates all commitments to detect false inputs, (iv) avoids online sampling operations (unlike lattice-based schemes) and random number derivations, all of which are error-prone, especially on low-end embedded devices, (v) it avoids using Forking Lemma, and therefore has tighter security reduction than traditional Schnorr-based signatures (with the aid of distributed verification process).

	\subsection{Performance on 8-bit AVR Microcontroller}
	
	\underline{\em Hardware and Software Configuration:} We fully implemented \lrsha~schemes on an 8-bit AVR ATmega2560 Micro-Controller Unit (MCU) at the signer side. This MCU is an 8-bit ATmega2560, having 256KB flash memory, 8KB SRAM, and 4KB EEPROM operating at a clock frequency of 16MHz. We used the $\mu$NaCl open-source software library to implement the EC-related operations and ASCON open-source software for cryptographic hash operations. 
	
	\begin{figure}[ht!]
		\centering
		\includegraphics[scale=0.52]{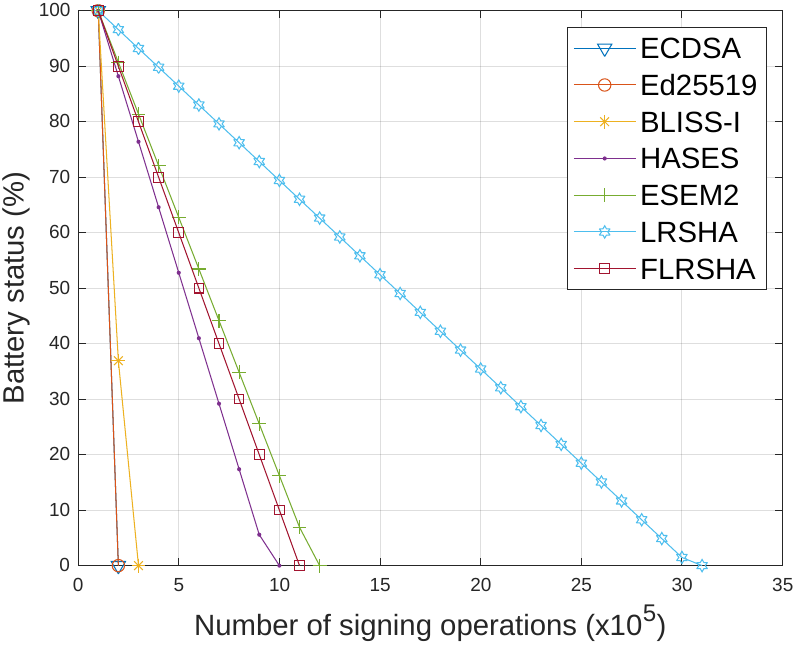}
		\caption{Impact of signing operations on the battery lifetime for \lrsha~and \flrsha~schemes and their counterparts}
		\label{fig:bench_signing}
	\end{figure}
	
	\noindent \underline{\em Performance Analysis:}
	Table \ref{tab:eses_8bit} compares the signing costs of our schemes with their counterparts on an 8-bit AVR MCU. 
	
$\bullet$ {\em Signing:} \lrsha~is $35.5\times$ and $3\times$ faster than the standard Ed25519 and ESEM, respectively. Table \ref{tab:signing_bench} showcases the signing efficiency of the most efficient candidate when pre-computation is considered. In our variant, the signer pre-computes symmetric keys via PRF calls to store them in memory offline and use them to generate signatures online. This strategy pushes the already efficient signing to the edge. For instance, the forward-secure \flrsha~with pre-computation is $8.5\times$ and $3\times$ faster than Ed25519-BPV and pre-computed $\text{ESEM}_2$, respectively, which are not FS.
	


	$\bullet$ {\em Energy Consumption:} Table \ref{tab:eses_8bit} depicts a comparative analysis of energy usage between \lrsha~and \flrsha~schemes with their respective counterparts on the selected MCU device. Specifically, we connected a pulse sensor to the 8-bit MCU. Then, we contrast the energy usage of a single sampling reading from the pulse sensor with that of a single signature generation. Our experiments validate that \lrsha~and \flrsha~exhibit superior performance when compared to the selected alternatives. This makes them the most suitable choices for deployment on resource-limited IoTs. 
	
	Figure \ref{fig:bench_signing} illustrates the impact of signing operations on an 8-bit MCU. Consistent with Table \ref{tab:eses_8bit}, it reaffirms the efficacy of our schemes in prolonging low-end device battery life. \lrsha~shows the longest battery life, depleting after $2^{30}$ signings. \flrsha, though slightly less efficient than $\text{ESEM}_2$, offers collusion resilience and authenticated decentralized verification without a single root of trust or key escrow, extending battery life beyond HASES.
	

\section{Conclusion} \label{sec:conclusion}
In this paper, we developed two new digital signatures referred to as {\em Lightweight and Resilient Signatures with Hardware Assistance} (\lrsha) and its forward-secure version (\flrsha). Our schemes harness the commitment separation technique to eliminate the burden of generation and transmission of commitments from signers and integrate it with a key evolution strategy to offer forward security. At the same time, they introduce hardware-assisted \comc~servers that permit an authenticated and breach-resilient construction of one-time commitments at the verifier without interacting with signer. We used Intel-SGX to realize distributed verification approach that mitigates the collusion concerns and reliance on semi-honest servers while avoiding single-point failures in centralized hardware-assisted signatures. Our distributed verification also permits offline construction of one-time commitments before the signature verification, thereby offering a fast online verification. 

Our new approaches translate into significant performance gains while enhancing breach resilience on both the signer and verifier sides. Specifically, to the best of our knowledge, \flrsha~is the only FS signature that has a comparable efficiency to a few symmetric MAC calls, with a compact signature size, but without putting a linear public key overhead or computation burden on the verifiers. Our signing process only relies on simple modular arithmetic operations and hash calls without online random number generation and therefore avoids complex arithmetics and operations that are shown to be prone to certain types of side-channel attacks, especially on low-end devices.  We formally prove the security of our schemes and validate their performance with full-fledged open-source implementations on both commodity hardware and 8-bit AVR microcontrollers. We believe that our findings will foster further innovation in securing IoT systems and contribute to the realization of a more secure and resilient IoT infrastructure.

	\printbibliography

\end{document}